\documentclass[journal=jacsat,manuscript=article]{achemso}

\usepackage{chemformula} 
\usepackage[T1]{fontenc} 

\usepackage{amsmath,amssymb}
\usepackage{graphicx}
\usepackage{booktabs}
\usepackage{siunitx}
\usepackage{tabularx}
\usepackage{geometry}
\usepackage{hyperref}
\hypersetup{
    colorlinks=true,
    linkcolor=blue,
    citecolor=blue,
    urlcolor=blue
}
\usepackage{soul}
\usepackage{xcolor}

\sethlcolor{pink} 

\author{Yalda Pedram}
\affiliation{Department of Mechanical and Materials Engineering, Queen’s University, Kingston, ON K7L 3N6, Canada}

\author{Yaoting Zhang}
\affiliation{Department of Mechanical and Materials Engineering, Queen’s University, Kingston, ON K7L 3N6, Canada}

\author{Scott Briggs}
\affiliation{Nuclear Waste Management Organization, Toronto M4T 2S3, Ontario, Canada}

\author{Chang Seok Kim}
\affiliation{Nuclear Waste Management Organization, Toronto M4T 2S3, Ontario, Canada}

\author{Laurent Brochard}
\affiliation{Laboratoire Navier, Ecole des Ponts ParisTech, Université Gustave Eiffel, CNRS, Marne-la-Vallée, France}

\author{Andrey G. Kalinichev}
\affiliation{Laboratoire SUBATECH, UMR 6457–Institut Mines Télécom Atlantique, Nantes Université, CNRS/IN2P3, 44307 Nantes, France}

\author{Laurent Karim Béland}
\email{laurent.beland@queensu.ca}
\affiliation{Department of Mechanical and Materials Engineering, Queen’s University, Kingston, ON K7L 3N6, Canada}

\title[An \textsf{achemso} demo]
  {Investigating the effect of Cu$^{2+}$ sorption in montmorillonite using density functional theory and molecular dynamics simulations}

\abbreviations{IR,NMR,UV}

\begin{document}







\begin{abstract}
Montmorillonite (MMT) is the main mineral component of bentonite, which is currently proposed as a sealing material in deep geological repositories (DGRs) for used nuclear fuel. In the Canadian program, which will utilize copper-cladded used fuel containers, safety analysis considers the effect of copper corrosion, during which Cu$^{2+}$ ions could potentially be adsorbed by the surrounding MMT. In such a scenario, ion exchange between Na$^+$ and Cu$^{2+}$ is expected. In this study, a multiscale approach that combines electronic density functional theory (DFT) and force-field-based molecular dynamics (MD) simulations was employed to study the effect of introducing Cu$^{2+}$ ions to MMT. An extension to the ClayFF force field is parametrized and validated using DFT to model how Cu$^{2+}$ interacts with clay systems. MD simulations were performed to calculate the interaction free energies between MMT platelets containing Cu$^{2+}$ ions (Cu-MMT) and compared them to inter-platelet interaction energies in Na-MMT and Ca-MMT. Our calculations suggest Cu-MMT develops swelling pressures between those of Ca-MMT and Na-MMT. Furthermore, our MD simulations suggest that Cu$^{2+}$ has MMT interlayer mobility that is significantly slower than that of Ca$^{2+}$.
\end{abstract}

\section{Introduction}
\label{sec:Introduction}
Bentonite is used by diverse industries \cite{guggenheim2006summary, murray2006bentonite} because of its high swelling potential, strong adsorption capacity, temperature durability, and low permeability \cite{clem1961industrial, abootalebi2018thermal, uddin2008clays}. It is added to drilling fluids within the oil and gas exploration sector \cite{bourgoyne1986applied, temraz2016mineralogy, afolabi2017properties}, acts as an impermeable barrier for geotechnical engineering projects to prevent water infiltration \cite{murray2006bentonite, carr1994industrial}, and is used in the ceramics industry as a binding agent \cite{murray2006bentonite}. In the pharmaceutical and medical domains, it is utilized as a drug delivery systems, as a binder in tablet formulations and for absorbing poisons \cite{park2016application, murray2006bentonite, robertson1986fullers}. In environmental engineering, bentonite is used as efficient adsorbent of heavy metals and contaminants during wastewater treatment processes \cite{alvarez2003removal,murray2006bentonite, liu2022molecular}.

Bentonite is also under consideration in the context of nuclear waste management, being proposed as a sealing material in deep geological repositories (DGRs) \cite{sellin2013use, murray2006bentonite, grambow2016geological}. Proposed DGRs involve multiple barriers, typically including a radioactive waste form enclosed within a long-lived container, surrounded by bentonite-based materials. In the Canadian design, the copper-coated canisters within this system will undergo changing environmental conditions, transitioning from an oxic environment over a few weeks or months to anoxic conditions thereafter \cite{giroud2018fate, psar}. During the initial repository stage, oxygen will interact with copper, resulting in a surface film of Cu$_2$O or CuO/Cu(OH)$_2$ \cite{aromaa2021oxidation}. Depending on groundwater chemistry, it is possible that some Cu ions would dissolve in the surrounding solution, likely in the form of Cu$^{2+}$ or possibly as an anionic complex such as \(\text{CuCl}_2^{-}\) in high chloride systems and would interact with the surrounding bentonite. This investigation focuses on the cationic species Cu$^{2+}$.

 
The primary component of bentonite is the montmorillonite (MMT) mineral. MMT consists of silicon tetrahedral (T) and aluminum/magnesium octahedral layers (O), forming TO and TOT layers at the nanoscale, with hydrated interlayers and hydroxyl group edges \cite{uddin2018montmorillonite, booker2004barrier}. A distinguishing feature of MMT is its permanent net negative charge, resulting from isomorphic metal substitution. For instance, Al$^{3+}$ or Fe$^{3+}$ can replace Si$^{4+}$ in T layers, while Mg$^{2+}$ or Li$^+$ can substitute Al$^{3+}$ in O layers \cite{booker2004barrier, odom1984smectite}. Layer-charged MMT is balanced by hydrated cations, such as Ca$^{2+}$ and Na$^+$. The swelling mechanism of these minerals is driven by the hydration of adsorbed ions between the layers due to their high hydration energy, leading to charge neutralization; such swelling clays are also known as smectite clays \cite{booker2004barrier, anderson2010clay, cygan2021advances}.

Several experiments have shown two fundamental processes governing MMT behavior: crystalline swelling and osmotic swelling \cite{norrish1954crystalline, norrish1954swelling, fink1964x}. 
Crystalline swelling, linked to hydrated interlayer cations, is stepwise and confined within a basal spacing range of $\sim$9-20 Å which is observed in a wide range of clay minerals \cite{norrish1954crystalline, anderson2010clay, zhang2022coarse}. In contrast, osmotic swelling is a continuous process, involving a range greater than $\sim$35-40 Å, and results from ion concentration disparities, also increases linearly with increasing water content \cite{norrish1954crystalline, anderson2010clay}. In the osmotic swelling range, Na-montmorillonite (Na-MMT) can absorb water, expanding significantly beyond the limits of crystalline swelling \cite{anderson2010clay}. Norrish \textit{et al.} \cite{norrish1954swelling} reported 20-fold volume expansion of Na-MMT in both osmotic and crystalline regions. This expansion is driven by a combination of electrostatic and Van der Waals forces, as well as pressure from cations between the clay layers \cite{norrish1954crystalline, norrish1954swelling}. Of note, the extent of swelling in Na-MMT is linked to multiple factors, including dry density, solution salinity, and temperature \cite{dixon2019review}.

There exists a transition region known as the 'Ravina and Low' zone \cite{ravina1977change}, which marks the abrupt shift between these two swelling regimes. However, this transition, referred to as \textit{forbidden} by Ravina and Low \cite{ravina1977change}, has not been observed in X-ray diffraction experiments \cite{zhang2014hydration, norrish1954swelling, fink1964x, foster1954lattice}. Fink \textit{et al.} \cite{fink1964x} demonstrated that this forbidden layer occurs in Wyoming Na-MMT, where isomorphic substitution occurs in both the tetrahedral and octahedral layers, and not in Na-MMTs with only octahedral substitutions. Furthermore, Meleshyn \textit{et al.} \cite{meleshyn2005gap} employed Monte Carlo simulations to reveal that interlayer spaces within such Na-MMTs can be locked due to a chainlike structure composed of substituted tetrahedron-Na cation-water molecule-Na cation-water molecule-Na cation-substituted tetrahedron' between layers. This space can remain locked until the water content approaches 850 mg/g.

Classical DLVO theory \cite{derjaguin1993theory}, named after its developers Derjaguin, Landau, Verwey, and Overbeek, describes the interactions between colloidal particles in a liquid medium. The theory considers two colloidal particles and characterizes their interaction as a balance between repulsive electrostatic forces and attractive Van der Waals forces. However, the DLVO theory fails to capture nanoscale forces between MMT platelets and the influence of divalent cations which makes it inadequate for understanding swelling behavior in the range of 1-2 nm \cite{pashley1984molecular, israelachvili1983molecular, jellander1988attractive}. Molecular dynamics (MD) simulations overcome these limitations and allow to better understand MMT at these very fine scales.  Atomistic MD simulations directly probe nanoscale interactions and are a good complement to traditional characterization methods such as scanning and transmission electron microscopy and X-ray diffraction (XRD) \cite{du2020revealing}.

ClayFF has been widely used in the literature for the atomistic simulations of clays and other mineral structures. Of particular interest for the present study, MMT is routinely modeled using the ClayFF force-field, developed by Randall T. Cygan and collaborators \cite{cygan2004molecular}. This force-field is based on electrostatic and Lennard-Jones pair interactions and will be elaborated upon in the method section. For example, Ho \textit{et al.} \cite{ho2019revealing} used a modified ClayFF force-field, as discussed later, to explore the impact of hydroxyl edges on Na-MMT hydration. Hydroxyl groups hindered the entry of the initial water layer (1-W) to the interlayer. The study indicated an alignment of the 0-W state at 9.85 Å, consistent with XRD results for dry MMT (9.6-9.8 Å) \cite{honorio2017hydration}. The study clarified the role of hydrogen bonds between hydroxyl layers 
as gatekeepers for water entry between interlayers. Sun \textit{et al.} \cite{sun2015estimation} used a ClayFF-based model to explore swelling pressure in semi-periodic Na-MMT sheets in various solutions; swelling pressure increased exponentially with dry density, regardless of the solution, with the highest swelling pressure observed in pure water. In another MD-based study, Akinwunmi \textit{et al.} \cite{akinwunmi2020swelling} examined the swelling characteristics of periodic Ca-MMT sheets and periodic Na-MMT sheets in pure water. According to the MD simulations, at low to moderate dry densities, Ca-MMT’s swelling pressure was lower than Na-MMT’s. At high dry density, Ca-MMT exerted greater swelling pressure than Na-MMT \cite{akinwunmi2020molecular}.

In order to study the influence of Cu$^{2+}$ on MMT at the atomistic level, existing ClayFF force-field need to be extended to handle Cu$^{2+}$ in clay. To do so, electronic density functional theory (DFT), which is an approximation to quantum mechanics, will be employed as a benchmark to fit this extended force-field. Our main objectives are twofold:
\begin{enumerate}
    \item Parameterize and validate ClayFF extension for Cu$^{2+}$: We aim to develop and validate an extension of ClayFF with new interaction parameters for Cu$^{2+}$ based on DFT calculations.
\end{enumerate}
\begin{enumerate}
    \item Investigate the comparative behavior of counter-ions in MMT: We seek to understand the interaction and behavior of MMT platelets with different counter-ions (Na$^{+}$, Ca$^{2+}$, and Cu$^{2+}$).
\end{enumerate}
To achieve these objectives, we performed MD simulations to calculate interaction free energies between MMT platelets containing Cu$^{2+}$ and Ca$^{2+}$ counter-ions and compared these to interaction energies with Na$^{+}$ counter-ions. In addition, we estimated the swelling pressure between MMT platelets and compared the cases of Na$^{+}$, Ca$^{2+}$, and Cu$^{2+}$. Finally, we computed the self-diffusion coefficient of these three cations when confined in the interlayer space.


This work is divided into the following sections: in section \texttt{\textbackslash Method} we will explain the computational procedures, simulation parameters and employed MMT models. Section \texttt{\textbackslash Results and analysis}  is divided into three sub-sections. First, we will discuss the results of fitting procedures, then the validation of the extended ClayFF force-field and finally the obtained PMFs, swelling pressures and mobility of cations will be shown.
\ref{sec:Method}
\section{Method}
\label{sec:Method}
ClayFF is not designed to handle Cu$^{2+}$ on MMT. To overcome this limitation we conducted a series of electronic density functional theory (DFT) simulations. We will begin by discussing the DFT setup. Following that, we will discuss the ClayFF force-field, which forms the basis of our simulations. The MD setup utilized to investigate the behavior of our systems will also be discussed, including an introduction to the potential of mean force (PMF) approach employed to calculate interaction free energies. 
 
\subsection{Density functional theory}
DFT simulations were performed as implemented in the open-source software Quantum Espresso \cite{giannozzi2009quantum} with projector-augmented waves (PAW) \cite{blochl1994projector, dal2014pseudopotentials, qe}
. The generalized gradient approximation (GGA) by Per\-dew--Bur\-ke--Ern\-zer\-hof (PBE) was employed to calculate the exchange-correlation energy \cite{perdew1996generalized}. We prepared a periodic Arizona MMT containing 2 × 1 × 1 unit cells with the size of 10.54 × 9.15 × 6.75 Å using the ATOM package \cite{holmboe2019atom} with a total charge of -2. To neutralize the negative charge of the MMT, we introduced one Cu$^{2+}$, resulting in Cu-MMT with the formula \([Cu].[(Si_{16}[Mg_2 Al_6]O_{48}(OH)_8]\). A similar formula was used for Ca-MMT by substituting Cu$^{2+}$ with Ca$^{2+}$. 
\newline In another scenario, similar structures were used in separate simulations with three water molecules, two Na$^{+}$, one Ca$^{2+}$, and one Cu$^{2+}$.

The aim of studying Ca-MMT was to provide a reference for comparison with Cu-MMT, due to their shared divalent cation nature. Additionally, the difference in ionic sizes between Ca$^{2+}$ and Cu$^{2+}$ provides insights into how cation size affects the swelling pressure and interaction energies in MMT. This comparative analysis was intended to improve the accuracy of MD calculations for Cu-MMT. The inclusion of Ca$^{2+}$ and Na$^{+}$ ions within the original ClayFF package facilitated a more precise evaluation, leveraging the established parametrizations for these ions.

Convergence tests were performed, revealing that a kinetic energy cut-off for wave functions of 1224.5 eV (90 Ry) for Cu-MMT and 680.28 eV (50 Ry) for Ca-MMT, along with a 2 × 2 × 2 Monkhorst–Pack k-point grid \cite{monkhorst1976special}, were sufficient. Next, geometric optimization of Cu- and Ca-MMT was performed. For this purpose, seven distinct Ca-MMT and twelve Cu-MMT configurations, differing in the Ca/Cu positions on the MMT surface, were constructed. The choice of configurations reflects the differing levels of prior research: since Ca-MMT has been extensively studied and its position within MMT is well understood, fewer configurations were necessary. In contrast, more configurations were used for Cu-MMT to thoroughly explore and find the most stable configuration A periodic simulation box with an initial dimension of 11 Å along the Z axis was employed. Volume and ion relaxation were run for each configuration using the Broyden–Fletcher–Goldfarb–Shanno (BFGS) algorithm \cite{fletcher2000practical}. The total energy and force convergence thresholds were set at 1×10$^{-4}$ eV and 1×10$^{-3}$ atomic units of force (where 1 atomic unit of force is \(e^2 / a_0^2\) ), respectively. In our DFT setup, we did not explicitly define the charge for the MMT and Ca/Cu. Instead, we used a single Cu or Ca atom. To determine the optimal configuration, the adsorption energy of the adsorbed atoms was computed using  Eq.~(\ref{eq:one}):
\begin{equation}
E_{ads} = E_{Ca/Cu-MMT} - (E_{Ca/Cu}-E_{MMT})
\label{eq:one}.
\end{equation}

Here, E$_{Ca/Cu-MMT}$ represents the energy of the Ca/Cu-MMT system after optimization, E$_{Ca/Cu}$ denotes the energy of a single Ca/Cu atom in a simulation box, and E$_{MMT}$ signifies the energy of the optimized MMT structure.


\subsection{Interatomic interaction potential}
This subsection introduces ClayFF, which serves as the foundation for our MD simulations. Then, we detail the process of extending ClayFF for Cu$^{2+}$ interactions.
\subsubsection{ClayFF}
ClayFF \cite{cygan2004molecular, cygan2021advances} is constructed using mostly
non-bonded interactions including Van der Waals in the form of Lennard-Jones (12-6) (L-J) function Eq.~(\ref{eq:two}) and Coulombic interaction energies Eq.~(\ref{eq:three}):
\begin{equation}
E_{VDW} = \sum_{i\neq j}4\epsilon_{ij} \left[ \left( \frac{\sigma_{ij}}{r_{ij}} \right)^{12} - \left( \frac{\sigma_{ij}}{r_{ij}} \right)^6 \right]
\label{eq:two}.
\end{equation}
\begin{equation}
E_{Coul} = \frac{e^2}{4\pi\epsilon_0} \sum_{i\neq j}\frac{q_i q_j}{r_{ij}}
\label{eq:three}.
\end{equation}
The parameters $\varepsilon_{ij}$ and $\sigma_{ij}$ denote the potential energy well depth and the distance at which the potential energy crosses 0. For simplicity, these parameters are presented for identical atoms, referred to as diagonal interaction terms. Off-diagonal terms are calculated by the Lorentz-Berthelot combination rules which use arithmetic and geometric means for $\sigma_{ij}$ and $\varepsilon_{ij}$ \cite{allen2017computer}, respectively.
ClayFF includes the flexible simple point charge (SPC) water model introduced by Berendsen \textit{et al.} \cite{berendsen1981interaction}. Within this model, each atom is represented by a partial charge at the center of the atom. Specifically, oxygen and hydrogen atoms within water molecules carry partial charges of -0.82 e and +0.41 e, respectively. While ClayFF maintains similar L-J parameters for all oxygen atoms as introduced in the SPC water model, the partial charges vary based on local coordination and environments. ClayFF was originally designed for infinite (periodic) clay sheets; however, several studies have modified ClayFF to investigate clay particles containing hydroxyl edges \cite{cygan2021advances, ho2019revealing, underwood2020large, dufresne2018atomistic, greathouse2009implementation, pouvreau2019structure, lammers2017molecular, pouvreau2017structure} by extending harmonic or Morse potential for O-H bonds, as well as harmonic potentials for the metal-O-H angles. In our present investigation, a harmonic potential is employed for the bonds and angles. 
\subsubsection{Training procedure for Cu$^{2+}$ interactions}
In order to train and validate a ClayFF Cu$^{2+}$ interaction model, molecular statics and MD simulations were conducted in the large-scale atomic/molecular massively parallel simulator (LAMMPS) software package \cite{plimpton1995fast, thompson2022lammps}, employing the modified ClayFF force-field \cite{cygan2021advances}. For all simulations, a cut-off radius of 9.5 Å along with Ewald summation to compute long-ranged electrostatic interactions in reciprocal space was applied.

To initiate the process, we employed a periodic 2 × 1 × 1 unit cell of MMT mimicking the setups utilized in DFT calculations. Within this unit cell, we considered twelve distinct Cu-MMT setups, mirroring those employed in DFT. For each of these unique Cu-MMT configurations, we performed energy minimization via a conjugate gradient algorithm \cite{hestenes1952methods} with at least 14 different sets of Cu$^{2+}$ L-J parameters sourced from existing literature \cite{majidi2022performance, moses2017evaluation, anitha2015removal, li2014taking, guan1996md, louw2022interaction, akiner2013prediction, khanmohammadi2019molecular, torras2013determination, li2013rational, liao2015development,duffour2004md, zhang2023molecular, knight2020interfacial}, along with 40 sets of L-J parameters in a range of $\varepsilon$ from 0.001 to 0.9 kcal/mol and $\sigma$ from 2.11 to 3.51 Å. This range was selected based on a set of L-J parameters found in the literature that gave us the closest structural similarity to the DFT results. Our aim here was to identify a set of L-J parameters that best aligned with the geometric outcomes obtained from DFT.

Having identified a set of parameters that provided geometry, specifically matching the d-spacing and Cu-O, Cu-Si distances, similar to our DFT results, we proceed with validation of the extended ClayFF.

First, we evaluated the energy of each configuration in two contexts:
\begin{enumerate}
    \item The energy difference of each configuration in ClayFF relative to the most stable structure identified in ClayFF.
    
    \item  The energy difference of each configuration in DFT relative to the most stable structure identified in DFT.
    
\end{enumerate}
We considered 12 configurations for Cu-MMT and 7 configurations for Ca-MMT by varying the positions of ions, and then calculated the energy differences to determine the stability of each configuration. To achieve this, we optimized each structure with both DFT and the extended ClayFF model. This allowed us to identify the local minima for each initial configuration and subsequently compare the energy of each configuration in relation to the lowest found energy in either DFT or ClayFF. This approach was implemented for both Ca- and Cu-MMTs to ensure a comprehensive assessment of the parameterization.

Second, we assessed the spatial arrangement of individual atoms within the Cu-MMT system as determined by ClayFF and compared it to that from DFT calculations. This validation procedure was applied to the Ca-MMT system. Our principal objective in this validation was to ascertain that the interatomic interactions involving Cu$^{2+}$ within the ClayFF force-field had achieved a level of consistency on par with that attained for Ca$^{2+}$ within the original ClayFF force-field.

In another validation test, since bentonite is a swelling clay, we conducted separate simulations with similar MMT structures containing three water molecules, one Ca$^{2+}$, one Cu$^{2+}$, and two Na$^{+}$ ions in the canonical (NVT) ensemble at 500 K using the Nose-Hoover thermostat. Forty configurations for each structure were extracted from the dump file and examined in SCF DFT calculations to compare the correlation between atomic forces generated by DFT, the original ClayFF for Na$^{+}$ and Ca$^{2+}$, and the extended ClayFF for Cu$^{2+}$.

Finally to investigate the behavior of aqueous Cu$^{2+}$, we initially created a large simulation box containing 2790 SPC water molecules with periodic boundary conditions in all dimensions. We simulated it in an isothermal-isobaric (NPT) ensemble at 300 K and 1 atm pressure. We assessed equilibration by monitoring density and the radial distribution function (RDF) of H-H, H-O, and O-O. The average density of equilibrated water during a 500 ps period was found to be 0.994 g/cm³, consistent with values from other sources \cite{izvekov2002car}. Afterward, one Cu$^{2+}$ was inserted in the middle of the simulation box, along with two Cl$^{-}$ placed at the farthest distance from Cu$^{2+}$ to minimize their interactions. The reason behind having two Cl$^{-}$ was that the system had to be neutral in order to employ the Ewald summation. Water molecules within a 2 Å radius of Cu$^{2+}$ were removed. Subsequently, the system underwent three separate simulations, each focusing on a single set of L-J parameter. Each system was pre-equilibrated in an NPT ensemble at 300 K and 1 atm for 100 ps, then equilibrated quantities were sampled in the canonical (NVT) ensemble at 300 K using Nose-Hoover thermostat for a duration of 500 ps. Validation of the new force fields parameters involved assessing geometric characteristics of the Cu$^{2+}$ ion and its surroundings in water, including RDF, coordination number (CN) and angular distribution function (ADF).

\subsection{Potentials of mean force}
To gain insights into the behavior of MMT in the presence of Na$^+$, Ca$^{2+}$, and Cu$^{2+}$ ions at the atomic scale, potentials of mean force (PMF) were computed. A PMF quantifies the variation in free energy between different states along a reaction coordinate, connecting the energies of the states of interest through a defined pathway \cite{trzesniak2007comparison, kastner2011umbrella, sprik1998free}. In our work, PMFs are derived using force constraint technique \cite{sprik1998free}, focusing on edge-to-edge and center-to-center MMT configurations. This technique involves sampling the forces acting on fixed reaction coordinates, as described by Eq.~(\ref{eq:four}):

\begin{equation}
\Delta A_{a-b} = \int_{r_a}^{r_b} F_r \, dr
\label{eq:four}.
\end{equation}  
Here, $\Delta A_{a-b}$ represents the difference in free energy between states 'a' and 'b', and it is computed by integrating the force required to maintain the MMT at a specified distance of $r_a$ and $r_b$.
\newline\ Three different models of MMTs were built using the ATOM package \cite{holmboe2019atom} to calculate PMFs.
\begin{enumerate}
    \item A small hexagonal MMT platelet which aligns with the methodology detailed in Zhang \textit{et al.} work \cite{zhang2022coarse}. The prior research by Underwood and Bourg emphasized the energetic favorability of hexagonal-shaped MMTs due to their minimized number of broken silicate rings \cite{lammers2017molecular, underwood2020large}. This model maintains a stoichiometric ratio of 4.25:1 for Al:Mg with the charge of -8. The structural formula for the Cu-MMT is
    \begin{multline*}
        [Cu_8].[Si_{168}.[Mg_{16} Al_{68}] O_{192} (OH)_{78}].
    \end{multline*}    
    \item A periodic sheet of MMT with the structural formula
    \begin{multline*}
        [Cu_{13}].[Si_{320}.[Mg_{26} Al_{134}] O_{800} (OH)_{160}].
    \end{multline*}    
    \item A semi-periodic sheet, non-periodic in the x and periodic in y direction, with the structural formula
    \begin{multline*}
        [Cu_{16}].[Si_{336}.[Mg_{28} Al_{140}] O_{824} (OH)_{200}]
    \end{multline*}
\end{enumerate}

In the small platelet and semi-periodic model hydroxyl groups were introduced to rectify the edges. Al-O-H and Si-O-H in the edges were integrated into the simulation via the extension of harmonic bonds and angles from the initial ClayFF force field \cite{pouvreau2019structure}.
Initially, a large box of water was structurally optimized using the conjugate gradient (CG) method. Subsequently, equilibration occurred under an NPT ensemble at a temperature of 300 K and a pressure of 1 atm to ensure proper thermal, pressure, and energy stabilization of the system. Following this, Na-MMT, Ca-MMT, and Cu-MMT were each introduced into the box of water through separate simulations. Each PMF graph was generated through a sequence of simulations, wherein the distances between two MMT platelets or sheets were incrementally increased by an average of 0.2 Å. In each scenario, water molecules within a 2 Å radius of each platelet were excluded to prevent overlapping water molecules with MMT platelets and to avoid undesired interactions in the initial steps of simulation. 
Across all simulations, metals in the octahedral layers remained fixed in all x, y, and z directions: 11 out of 42 for platelets, 12 out of 160 for periodic sheets, and 14 out of 168 for semi-periodic sheets.

Subsequently, every system underwent pre-equilibration under NPT conditions at 300 K and 1 atm for 50 ps, followed by further equilibration and force sampling under NVT conditions at 300 K for 1 ns. A timestep of 1 fs was chosen for all simulations.

Swelling pressure calculations for Na-MMT, Ca-MMT, and Cu-MMT were determined by analyzing the gradients of PMFs at various dry densities. These dry densities, defined as \( \frac{m}{A \times d} \), where \( m \) is the mass of the clay with ions, \( A \) is the area of the clay, and \( d \) is the d-spacing, were calculated based on the center-to-center distances between two cation-MMTs, as inspired by the approach detailed in Sun's study \cite{sun2015estimation}. In addition to this analysis, we also computed the mean square displacement (MSD) for each counter-ion within the MMT clays.

\section{Results and analysis}
\label{section:Results and analysis}
\subsection{\label{sec:citeref}$\text{Cu}^{2+}$ interatomic interactions model}
\subsubsection{Results of the fitting procedure}
\begin{figure}[htbp]
    \centering
    \includegraphics[width=0.8\textwidth]{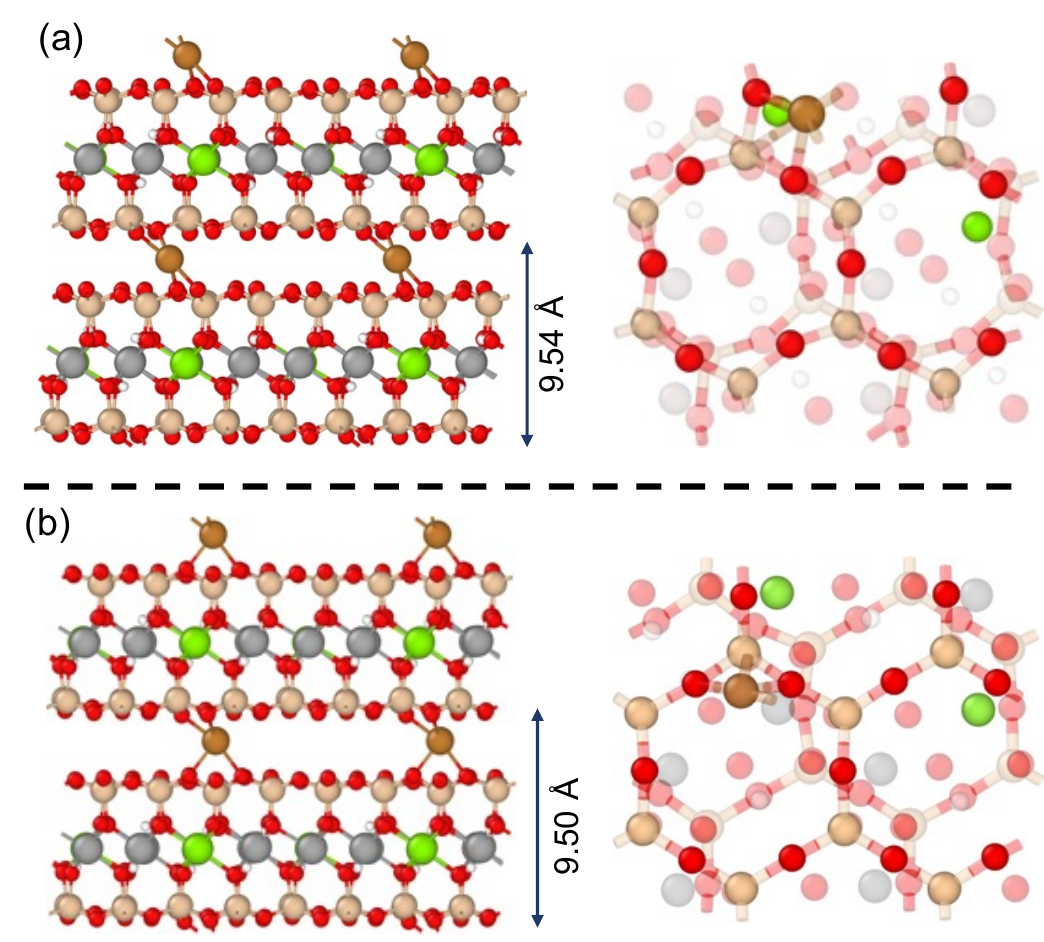}
    \caption{Structural comparison between (a) DFT-optimized periodic Cu-MMT and (b) Cu-MMT optimized using the extended ClayFF with $\sigma$  = 2.21 Å and $\varepsilon$ = 0.05 kcal/mol. The arrows indicate the d-spacing values. In this representation, brown, red, green, gray and beige spheres represent Cu$^{2+}$ ions, oxygen, magnesium, aluminum, and silicon atoms, respectively. The right side figures show a top view of the structure with the bottom tetrahedral and octahedral layers made transparent for better illustration.} 

    \label{fig:fig1}
\end{figure}

We start by presenting our DFT-calculated adsorption energies, providing an insight into the strength of interaction between cations and MMT surface. An adsorption energy of -255.73 kcal/mol for Cu$^{2+}$ cations was obtained. The optimal position for Cu$^{2+}$ was found just above the MMT surface, nearly in the middle of interlayer region, positioned close to the boundary of the ring, directly above one of the two Al$\rightarrow$Mg substitutions in the octahedral layer. Cu$^{2+}$ exhibited a bonding distance of 2.75 Å with Si atom and 2.08 Å and 2.06 Å with O atoms surrounding the Si atom in the tetrahedral layer of MMT. Details provided in Table~\ref{tab:table1}, and Fig~\ref{fig:fig1} illustrates the corresponding Cu-MMT structure. An important observation here is the distinct behavior of Cu$^{2+}$ compared to Ca$^{2+}$ in terms of adsorption on MMT surfaces. The adsorption energy for Ca$^{2+}$ for the corresponding configuration shown in Fig~\ref{fig:fig1} was determined to be -147.82 kcal/mol. Interestingly, in the case of Ca$^{2+}$, the ion is positioned precisely in the center of a ring and in close proximity to one of the MMT surfaces, consistent with prior calculations \cite{voora2011density}. By comparing the structures shown in Fig~\ref{fig:fig1}, we can understand the differences in adsorption characteristics between Cu$^{2+}$ and Ca$^{2+}$.

\begin{table*}[htbp]
    \caption{Structural details of optimized Cu-MMT in DFT and our best-fitted L-J parameters ($\sigma$ = 2.13 (Å) and $\varepsilon$ = 0.05 kcal/mol) for Cu$^{2+}$ in ClayFF.}
    \centering
    \begin{tabular}{lccc}
        \hline
        \textrm{Method} & \textrm{d-spacing (Å)} & \textrm{Cu-Si distance (Å)} & \textrm{Cu-O distance (Å)} \\
        \hline
        DFT & 9.54 & 2.75 & 2.08, 2.06 \\
        ClayFF ($\sigma$ = 2.13 and $\varepsilon$ = 0.05) & 9.35 & 2.67 & 2.02, 2.02 \\
        ClayFF ($\sigma$ = 2.16 and $\varepsilon$ = 0.05) & 9.50 & 2.66 & 2.05, 2.02 \\
        ClayFF ($\sigma$ = 2.21 and $\varepsilon$ = 0.05) & 9.50 & 2.68 & 2.06, 2.05 \\
        \hline
    \end{tabular}
    \label{tab:table1}
\end{table*}

After identifying the most stable Cu-MMT configuration using DFT, we aimed to replicate this structure with ClayFF. We performed a search involving 14 parameter sets to achieve accurate d-spacing, Cu-Si and Cu-O distances. The best-fitting L-J parameters were determined to be $\sigma$ = 2.13 Å and $\varepsilon$ = 0.05 kcal/mol, as were proposed by K. Anitha \textit{et al.} \cite{anitha2015removal}. To further enhance the structural accuracy, we explored variations in $\sigma$, including $\sigma$ = 2.16 and 2.21 Å, as detailed in Table~\ref{tab:table1}. $\sigma$ = 2.21 Å resulted in a structure closely resembling the DFT results. Fig~\ref{fig:fig1}(b) illustrates the optimized Cu-MMT structure associated with the extended ClayFF. Note that Cu$^{2+}$ as described by ClayFF resides in a different ring as compared to DFT, while maintaining similar bonding lengths and d-spacing as observed in the DFT simulations. In an effort to validate the positioning of Cu$^{2+}$ ions, we attempted to place the Cu$^{2+}$ ion near the DFT-derived position and subsequently relaxed the structure. However, during the relaxation process, the Cu$^{2+}$ ion consistently migrated to another ring and did not remain at the initial DFT position.

In addition to the parameter sets explored in this study, we evaluated a set of parameters reported by Zhang \textit{et al.} in a recent paper \cite{zhang2023molecular}: $\sigma$ = 3.114 Å and $\varepsilon$ = 0.005 kcal/mol. This L-J model closely approximated our findings, with Cu-Si distances at 2.67 Å and Cu-O distances at 2.14 Å and 2.09 Å, along with a d-spacing of 9.64 Å. While these parameters effectively described the Cu-MMT system in a dry environment, similar to our DFT results, our developed model exhibited higher accuracy. It is worth noting that the value for $\varepsilon$ is expressed in a different unit compared to the cited value in reference \cite{zhang2023molecular}, but they are equivalent.

\begin{figure}[htbp]
\centering
\includegraphics[width=0.7\textwidth]{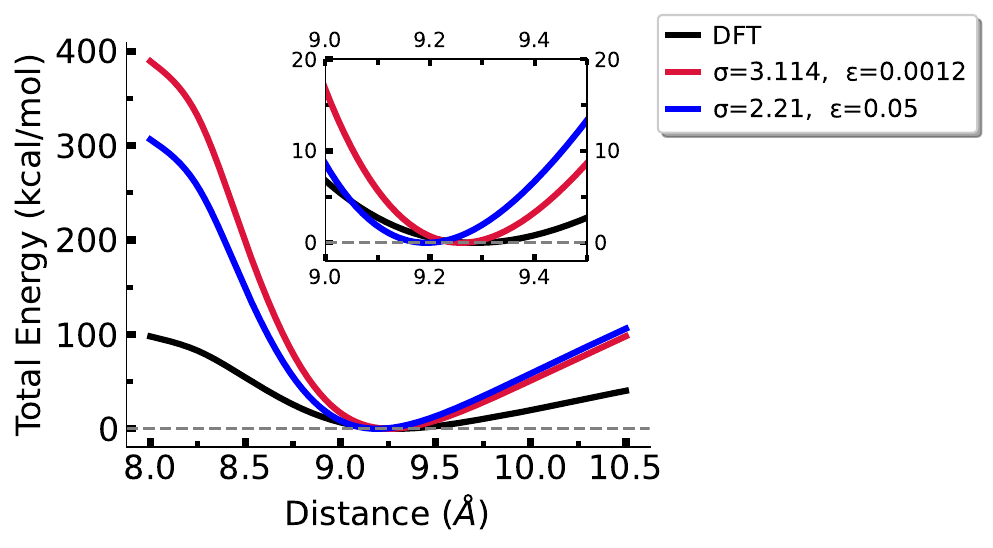}

\caption{\label{fig:fig2} A comparative analysis of the Cu-MMT energy profile along the perpendicular direction of MMT surfaces. The energy profiles are generated from DFT (black line), our best-fitted L-J parameters (blue line), and the parameters proposed by Zhang \textit{et al.} (red line) \cite{zhang2023molecular}.
}
\end{figure}

We conducted a series of calculations to assess how Cu-MMT energies vary as Cu$^{2+}$ traverses the interlayer space. This involved testing 40 sets of L-J parameters ranging from $\sigma$ = 2.11 to 2.91 Å and $\varepsilon$ = 0.001 to 1 kcal/mol, in addition to the parameters employed by Zhang \textit{et al.} \cite{zhang2023molecular}. The results, illustrated in Fig~\ref{fig:fig2}, demonstrate the outcomes obtained using Zhang's parameter set. 

The results, illustrated in Fig~\ref{fig:fig2}, demonstrate the outcomes obtained using Zhang's parameter set compared with our results and the DFT results. The zoomed-in graph highlights the changes in energy around the minimum well, where the probability of finding Cu$^{2+}$ is highest. Even under extreme stresses, the Cu$^{2+}$ ion remains within the range of 9 to 9.5 Å from its equilibrium position. Energy curves beyond this range do not significantly impact the mechanical properties due to the large energy scale compared to thermal agitation energy (kT). Our model is similar to Zhang's, both depicting similar potential energy changes at the thermal scale with comparable well depths. The blue line matches the DFT results around the minimum as accurately as Zhang's model, indicating similar interaction characteristics.



\subsubsection{Validation}

\begin{figure*}[htbp]
\centering
\includegraphics[width=1\textwidth]{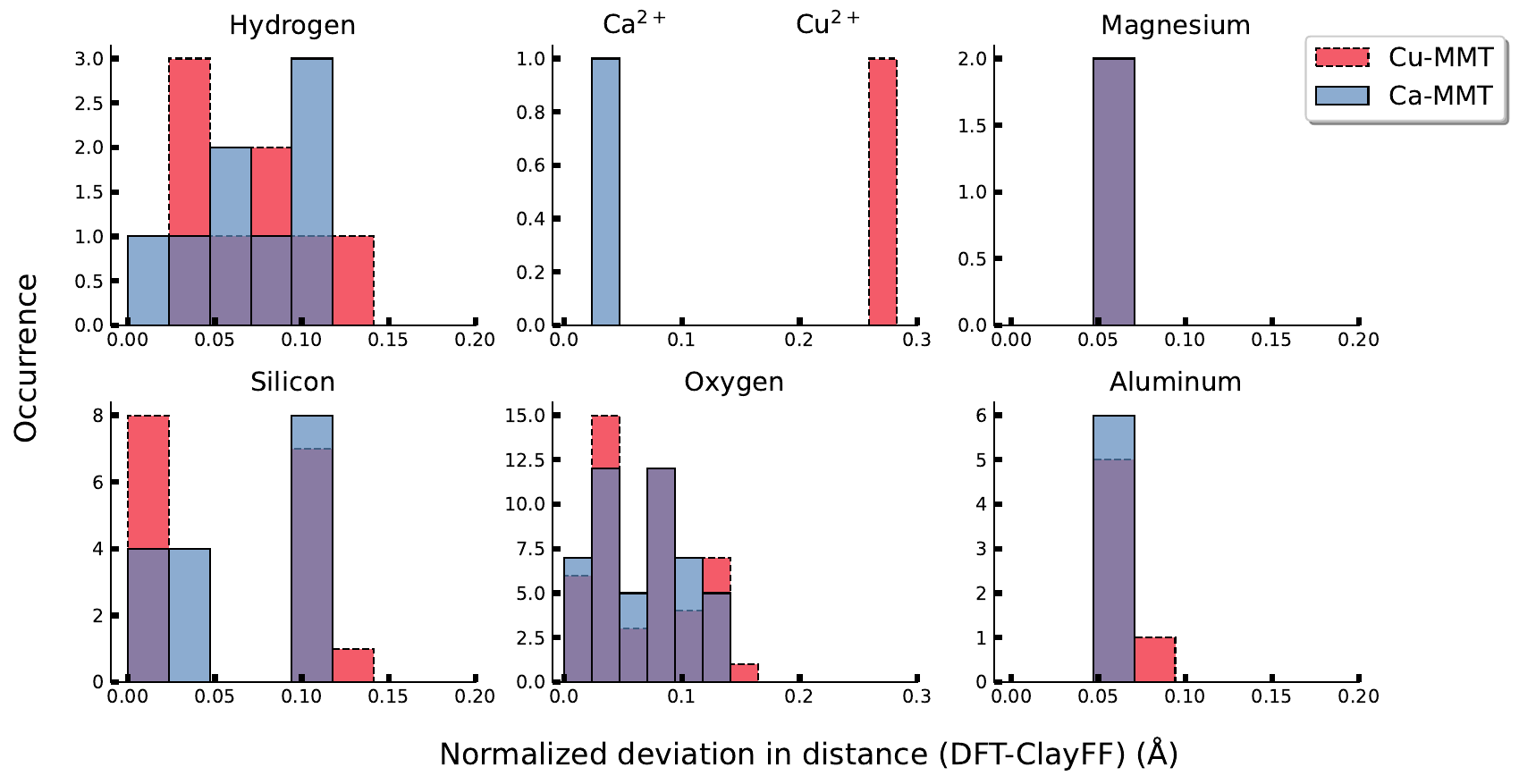}
\caption{\label{fig:fig3} Comparison of particle positions as predicted by ClayFF and DFT after structural relaxation. The distances are normalized to the unit cell dimensions. Ca-MMT and Cu-MMT deviations are shown in blue and red, respectively, with overlapping deviations displayed in purple.
}
\end{figure*}

First, the positions of each element in the ClayFF-optimized Cu-MMT and their counterparts in DFT simulations were compared.
The atomic coordinates from both simulations were converted to fractional space for accurate comparison. One atom was used as a reference in both the DFT and ClayFF simulations, and the positions of all other atoms were normalized relative to this reference atom. The deviations in the position of each element between ClayFF and DFT, normalized to the unit cell dimensions, are visually presented in Fig~\ref{fig:fig3}. We also optimized Ca-MMT using both DFT and ClayFF and quantified the deviations in atomic positions from the DFT-optimized structure. The extended Cu$^{2+}$ ClayFF model led to deviations for the MMT atoms comparable to those associated with the Ca$^{2+}$ interaction model in ClayFF inside MMT. Although the deviation for Cu$^{2+}$ is higher than for Ca$^{2+}$, the bonding length and d-spacing are similar to DFT. This higher deviation for Cu$^{2+}$ can be attributed to the movement of Cu to another ring, which caused relatively high deviation for a small cell size.

\begin{figure}[htbp]
    \centering
    \includegraphics[width=0.8\textwidth]{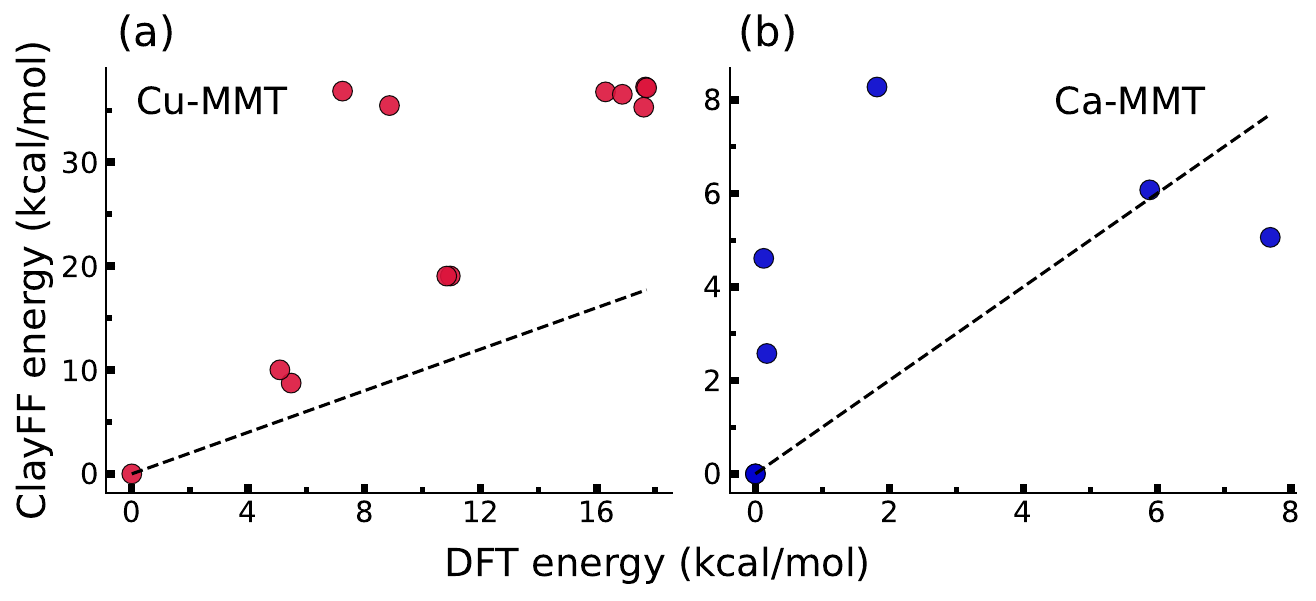}
    \caption{Adsorption energy--setting the lowest-energy site at zero--comparison. ClayFF and DFT calculations describing (a) Cu$^{2+}$ and (b) Ca$^{2+}$ adsorbed on different MMT surface sites after structural relaxation are compared. Dashed lines represent a one-to-one relationship.}
    \label{fig:fig4}
\end{figure}

Another analysis involved comparing the Ca$^{2+}$ and Cu$^{2+}$ adsorption energies for each configuration relative to the minimum energy state using either ClayFF or DFT. For this analysis, we optimized each configuration to find the local minimum around the initial configuration and compared the obtained energies. In Fig~\ref{fig:fig4}, each point corresponds to an initial configuration. As illustrated, the extended ClayFF successfully described the lowest-energy Cu$^{2+}$ site. Notably, while similar initial configurations led to the lowest energy for both DFT and ClayFF, in ClayFF, Cu$^{2+}$ moves to a different ring than in DFT. Specifically, in DFT optimization, Cu$^{2+}$ positions itself above the Al$\rightarrow$Mg substitutions in the octahedral layer, whereas in ClayFF, the ion moves above an Al atom in a ring without substitution. However, the distances between Cu$^{2+}$ and O and Si atoms in the tetrahedral layer of MMT are very similar in both methods. The DFT-ClayFF correlation pertaining to Cu-MMT is more robust than that pertaining to Ca-MMT. According to DFT, the variation of the adsorption energies in Cu-MMT are larger than those involved in Ca-MMT. ClayFF captures this effect, albeit exaggeratedly so. DFT suggests Cu$^{2+}$ adsorption energies range between 0 and 17 kcal/mol--relative to the lowest-energy site--ClayFF suggest a 0 to 35 kcal/mol range, as compared to 0 to 8 kcal/mol for Ca-MMT. Notably, low-energy Cu-MMT configurations, which are of higher importance, show relatively good correlation between DFT and ClayFF. However, low-energy configurations of Ca-MMT in DFT (energies around 0.1 kcal/mol) are not well predicted by ClayFF.

\begin{figure}[htbp]
    \centering
    \includegraphics[width=1\textwidth]{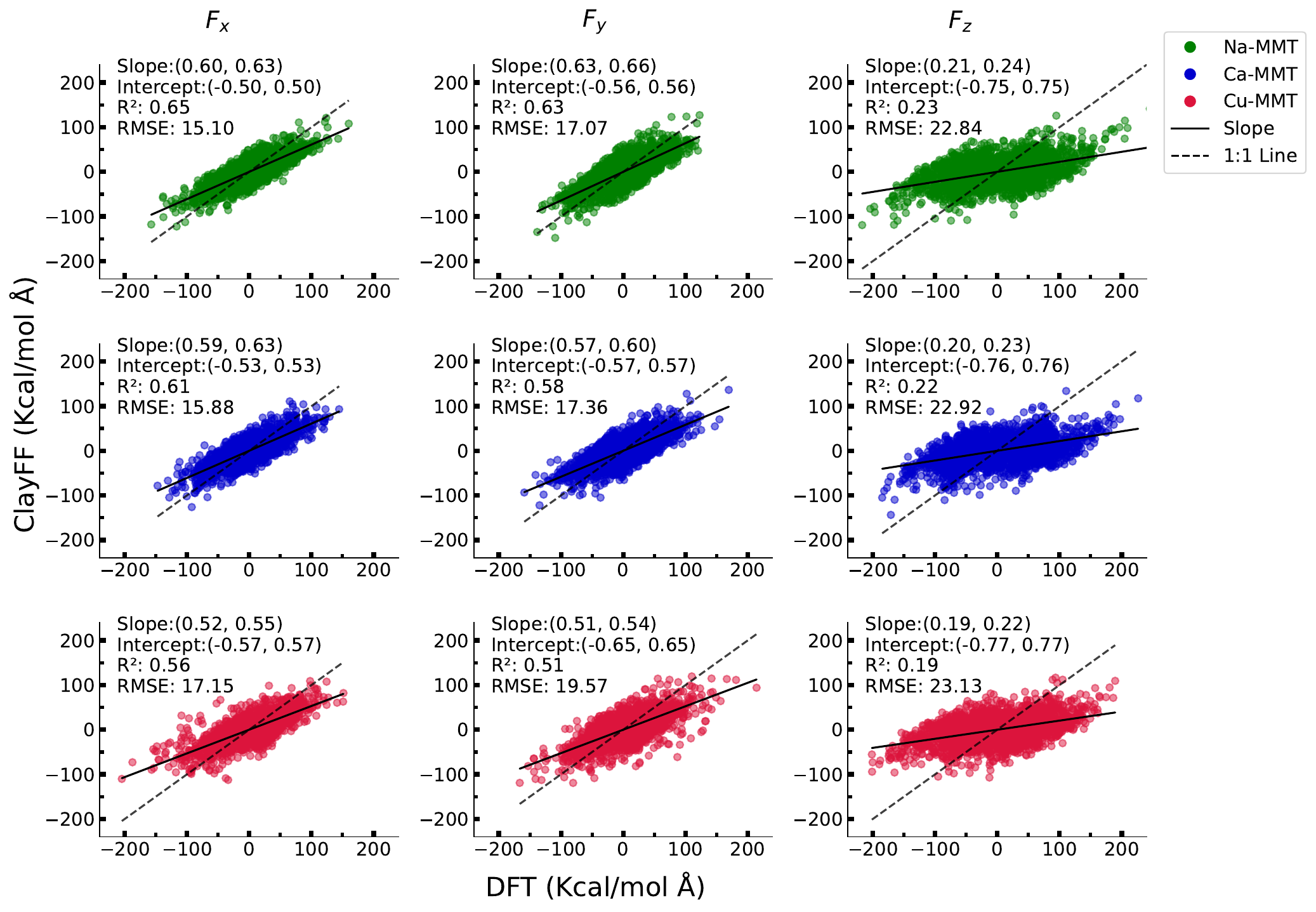}
    \caption{Comparison of atomic forces on atoms inside the Na- (green), Ca-(blue), and Cu-MMT (red) systems, calculated using SCF DFT (x-axis) and ClayFF (y-axis). Each row corresponds to a different ion type, while the columns represent forces in the x, y, and z directions respectively. The slope, intercept, their 95\% confidence interval, R$^2$ value, and RMSE are provided for each fit. The solid line indicates the linear regression fit, and the dashed line represents the ideal 1:1 correlation.}
    \label{fig:fig5}
\end{figure}

\begin{figure}[htbp]
    \centering
    \includegraphics[width=1\textwidth]{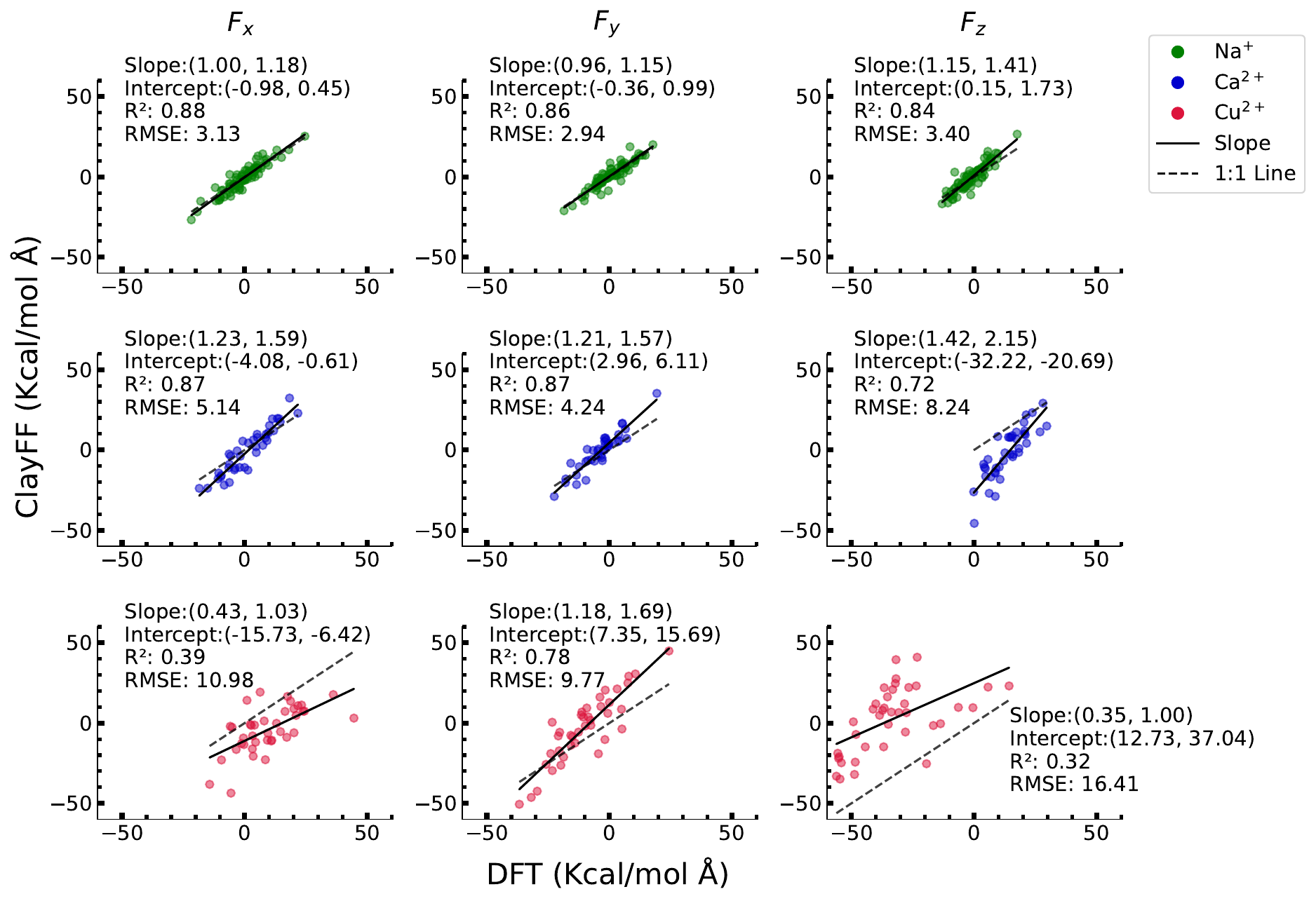}
    \caption{Comparison of atomic forces on ions in the Na- (green), Ca- (blue), and Cu-MMT (red) systems, calculated using SCF DFT (x-axis) and ClayFF (y-axis). Each row corresponds to a different ion type, while the columns represent forces in the x, y, and z directions, respectively. The slope, intercept, their 95\% confidence interval, R$^2$ value, and RMSE are provided for each fit. The solid line indicates the linear regression fit, and the dashed line represents the ideal 1:1 correlation.}
    \label{fig:fig6}
\end{figure}

To validate the force field, we conducted NVT ensemble simulations at 500 K for Na-, Ca-, and Cu-MMT structures with three water molecules using ClayFF. We extracted forty configurations for each structure and performed SCF DFT calculations to compare the correlation between atomic forces generated by DFT and ClayFF (original for Na$^{+}$ and Ca$^{2+}$, and extended for Cu$^{2+}$). The resulting forces on atoms inside MMT, including water molecules, are shown in Fig~\ref{fig:fig5}, and the forces on the counter-ions are shown in Fig~\ref{fig:fig6}. Each point represents a force calculated by SCF DFT (x-axis) and ClayFF (y-axis). The solid line indicates the linear regression fit, while the dashed line represents the ideal 1:1 correlation.
Fig~\ref{fig:fig5} shows the comparison of atomic forces on atoms inside the Na- (green), Ca- (blue), and Cu-MMT (red) systems. For the Na-MMT system, the data points are closely clustered around the 1:1 line, indicating strong agreement between DFT and ClayFF forces. The high R$^2$ values and low RMSE demonstrate the accuracy of the original ClayFF for Na$^{+}$ ions. In the Ca-MMT system, which is included in the original ClayFF, the results show good agreement with DFT, albeit with higher RMSE and lower R$^2$ values compared to Na-MMT. This reflects the inherent reliability of ClayFF for Ca$^{2+}$ ions. Interestingly, the Cu-MMT system exhibits statistical trends similar to the Ca-MMT system. The slopes and intercepts for Cu-MMT are close to those of Ca-MMT, with comparable R$^2$ values and RMSE. This suggests that the extended ClayFF for Cu$^{2+}$ performs reasonably well, achieving a level of accuracy similar to Ca$^{2+}$, which is originally parameterized in ClayFF.

In Fig~\ref{fig:fig6}, we observe the comparison of atomic forces on ions in the Na- (green), Ca- (blue), and Cu-MMT (red) systems. For the Na-MMT system, the forces on Na$^{+}$ ions show a high degree of correlation with the DFT results, evidenced by slopes and intercepts close to the ideal values, high R$^2$ values, and low RMSE. This indicates a strong agreement between ClayFF and DFT forces. In the Ca-MMT system, the comparison shows good agreement with DFT forces for Ca$^{2+}$ ions. The slopes and intercepts, although slightly different from the ideal values, still indicate a high correlation, with reasonably high R$^2$ values and low RMSE, though not as strong as in the Na$^+$ system. For the Cu$^{2+}$ ions, while there is more deviation compared to Na$^{+}$, the statistical measures (slopes, intercepts, R$^2$ values, and RMSE) in the y-direction are similar to those of the Ca$^{2+}$ ions, and results in other directions show similar accuracy to those for atoms inside Ca-MMT. It should be noted that Cu$^{2+}$'s partially filled d-orbitals lead to more complex chemical properties and structures compared to Ca$^{2+}$ and Na$^{+}$. Consequently, performing classical force field calculations for Cu$^{2+}$, such as the simple L-J interaction energies used in ClayFF, does not achieve the same level of accuracy as for simpler ions like Ca$^{2+}$ and Na$^{+}$.

\begin{figure}[htbp]
    \centering
    \includegraphics[width=0.8\textwidth]{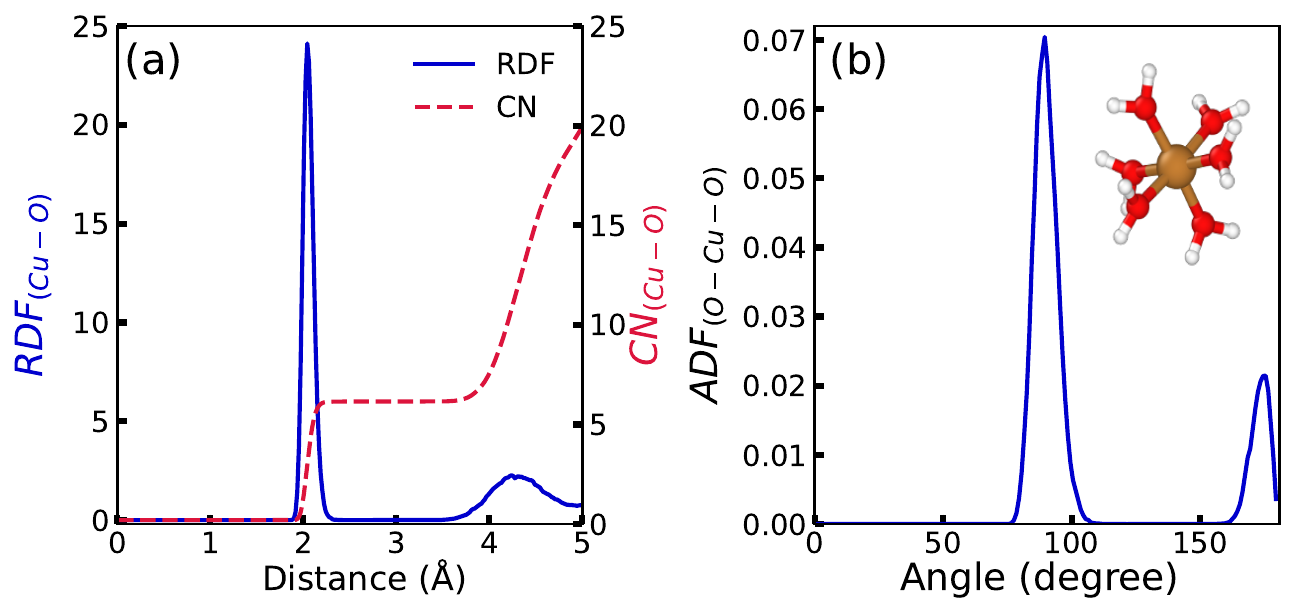}
    \caption{(a) Radial distribution function (RDF) of Cu-O$_{Water}$ for hydrated Cu$^{2+}$ in the flexible SPC water model. The solid blue line represents the RDF and the dashed red line represents the coordination numbers (CN). (b) Angular distribution function (ADF) of O$_{Water}$-Cu-O$_{Water}$ in the first hydration shell.
}
    \label{fig:fig7}
\end{figure}

Validation of the extended ClayFF model in a hydrated environment involved assessing the geometric properties of Cu$^{2+}$ ions immersed in water, including RDF, CN and ADF. We considered the three best-performing sets of  L-J parameters found as detailed in Table~\ref{tab:table2}. Fig~\ref{fig:fig7} highlights the results achieved using the optimal parameters, $\sigma$ = 2.21 Å and $\varepsilon$ = 0.05 kcal/mol. In Fig~\ref{fig:fig7}, the RDF for Cu-O$_{Water}$ is depicted by the blue line, while the hydration number of Cu$^{2+}$ is shown in red. The first peak in the RDF corresponds to the average distance between Cu$^{2+}$ and the oxygen atoms in its first hydration shell, measuring 2.04 Å. A visual representation of Cu$^{2+}$'s first hydration shell is provided in Fig~\ref{fig:fig7}. The second peak in the RDF, located at 4.24 Å, signifies the distance between Cu$^{2+}$ and the oxygens in the second hydration shell. Coordination numbers were determined by integrating the Cu-O$_{Water}$ RDF. The analysis of the first coordination number, representing the count of oxygen atoms in the first hydration shell, confirms that the new force field favors a 6-fold coordination for Cu$^{2+}$ cations in the water solution, consistent with previous studies, as elaborated in Table~\ref{tab:table2} \cite{marini1999investigation, bowron2013hydration, schwenk2003extended, van2010development, ohtaki1993structure}. The second-shell coordination number, calculated for distances between 3 Å and 4.8 Å, was found to be 12.3, closely aligning with results from quantum mechanical/molecular mechanical molecular dynamics (QM/MM MD) simulations \cite{schwenk2003extended}, which are also compared in Table~\ref{tab:table2}. Additionally, the ADF of O-Cu-O for the first hydration shell around the Cu$^{2+}$ ion, as presented in Fig~\ref{fig:fig7}, indicated angles of 89.5° and 175.5°. These results demonstrate excellent agreement with findings from other studies \cite{schwenk2003extended, amira2005distorted, van2010development, ohtaki1993structure, magini1982coordination, musinu1983coordination}. However, note that certain quantum effects observed using QM approaches, such as Jahn-Teller inversion, cannot be replicated using classical force-fields.

\begin{table*}[t]
    \centering
    \scriptsize 
    \caption{Comparison of Cu$^{2+}$ hydration shell properties obtained using various methodologies, including Cu-O$_{\text{Water}}$ distances and coordination numbers (CN) for the first and second hydration shells, along with the O-Cu-O angle in the first hydration shell.}
    \begin{tabular}{l p{4cm} p{1.2cm} p{2cm} p{1cm} p{2cm}}
        \hline
        Method  & Cu-O$_{\text{1st shell}}$ (Å) & CN$_{\text{1st}}$ & Cu-O$_{\text{2nd shell}}$ (Å) & CN$_{\text{2nd}}$ & Angle$_{\text{O-Cu-O}}$ (°) \\  
        \hline
        ClayFF ($\sigma$ = 2.13, $\varepsilon$ = 0.05) & 2.02 & 6 & 4.28 & 12.8 & 88.5, 175.5 \\
        ClayFF ($\sigma$ = 2.16, $\varepsilon$ = 0.05) & 2.02 & 6 & 4.27 & 12.64 & 88.5, 174.5 \\
        ClayFF ($\sigma$ = 2.21, $\varepsilon$ = 0.05) & 2.04 & 6 & 4.24 & 12.3 & 89.5, 175.5 \\
        HF/MM+3bd MD \cite{marini1999investigation} & 2.08 & 6 & 4.2 & 14.5 &  \\
        EXAFS/LAXS \cite{bowron2013hydration} & 1.95, 2.29 & 6 & 4.17 & 8 &  \\
        B3LYP QM/MM \cite{schwenk2003extended} & 2.02, 2.29 & 6 & 4.13 & 11.9 & 89, 173 \\
        CPMD \cite{pasquarello2001first} & 1.96, 1.96 & 5 &  &  &  \\
        CPMD \cite{amira2005distorted} & 2, 2.45 & 5 & 4.03 & 8 & 90, 180 \\
        ReaxFF \cite{van2010development} & 1.94, 2.27 & 6 & 4.27 & 12.5 & 90, 176 \\
        Neutron diffraction \cite{ohtaki1993structure, beagley1989computational} & 1.97 & 6 &  &  &  \\
        X-ray diffraction \cite{ohtaki1993structure, magini1982coordination} & 1.98, 2.39 & 6 & 3.96 & 11.1  &  \\
        X-ray diffraction \cite{ohtaki1993structure, magini1982coordination} & 1.98, 2.34 & 6 & 3.95 & 11.6 &  \\
        X-ray diffraction \cite{ohtaki1993structure, musinu1983coordination} & 2.01, 2.33 & 6 & 4.2 & 7.6 &  \\
        \hline
    \end{tabular}
    \label{tab:table2}
\end{table*}

\subsection{\label{sec:citeref}Potentials of mean force}
We conducted a series of MD simulations using the extended ClayFF force field to calculate potentials of mean force (PMF) profiles in wet MMT systems. Our study includes four different scenarios, including both center-to-center and edge-to-edge configurations of MMT platelets. The rationale behind exploring various systems lies in the recognition that MMT platelets in real-world settings exhibit a wide range of sizes, shapes, and spatial arrangements when interacting with each other.
\subsubsection{Measured potentials of mean force}
\begin{figure*}[htbp]
    \centering
    \includegraphics[width=1\textwidth]{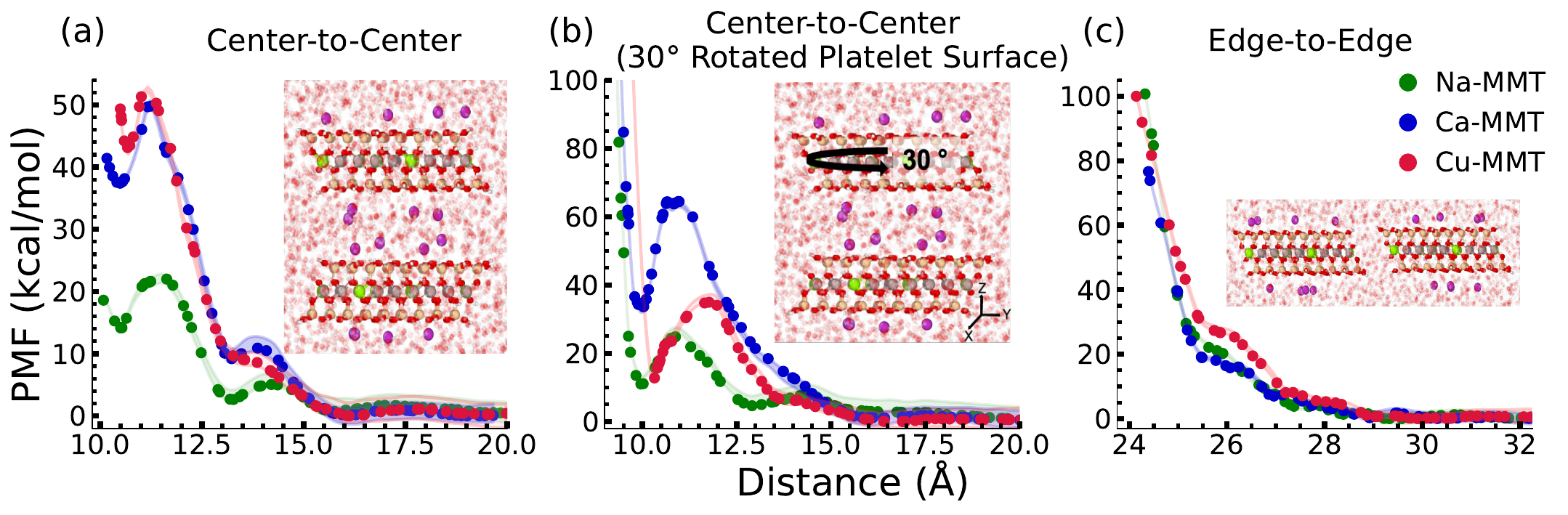}
    \caption{Potentials of mean force (PMF) for (a) center-to-center interactions of Na-, Ca-, and Cu-MMT platelets, (b) center-to-center interactions with a 30° rotation applied to one of the platelet surfaces and (c) edge-to-edge interactions of the aforementioned structures. The PMF results, represented by dots, are averaged, and the shaded colors indicate the variability obtained from 100 bootstrap samples.
}
    \label{fig:fig8}
\end{figure*}

Our investigation was initiated by examining the behavior of two center-to-center MMT platelets, including Na$^+$, Ca$^{2+}$, and Cu$^{2+}$, in separate simulations. The resulting average PMF curves, along with a snapshot illustrating the Na-MMT structure at a specific center-to-center separation, are presented in Fig~\ref{fig:fig8}.
Remarkably, the PMF curves, irrespective of the cation type, reveal a rich landscape of local energy minima, corresponding to various hydration states. These states encompass configurations with zero water layers (0-W), one water layer (1-W), and two water layers (2-W) inserted between the MMT platelets. Notably, the transitions between the 0-W and 1-W states, as well as between the 1-W and 2-W states, consistently occur at distances of approximately 2.5-3 Å. This distance closely matches the diameter of a water molecule and aligns with experimental measurements of hydration forces observed between diverse surfaces, including mica and MMT materials \cite{israelachvili1983molecular, clark1937study, antognozzi2001observation, pashley1984molecular}.
There is a pronounced repulsive interaction at distances less than 16 Å. There are substantial energy barriers situated at 10.5-11.5 Å for the hydration process from the 0-W to 1-W state. These energy barriers can be partly attributed to the contribution of hydrogen bonding at the edges of the platelets. This observation is consistent with a prior study \cite{ho2019revealing}, which demonstrated that hydrogen bonds at the platelet edges serve as gatekeepers, controlling the entry of water molecules into the interlayer spaces of MMT. This effect primarily influences the first layer of hydration. 
The initiation of the swelling process is found to depend on breaking hydrogen bonds and was influenced by the size and shape of the MMT platelets \cite{ho2019revealing}. The differences between the two local energy minima indicate the relative stability of different hydrated states. The significant difference between minima at 0-W and 1-W for all structures underscores the strong preference of MMT platelets for water absorption. This behavior is expected for MMT in contact with liquid water under standard conditions, as opposed to low hydration states which become more favorable in vapor when relative humidity (RH) is reduced. Furthermore, the energy difference between 0-W and 1-W of Na-MMT is lower than that observed for Ca- and Cu-MMT, possibly due to variations in the hydration energy of the first water shell around different cations. Additionally, the larger size of hydrated Cu$^{2+}$ leads to its local energy minima being located at a slightly greater distance as compared to the other two cations.

Additionally, to address concerns about the anisotropy of platelets and their rotational degrees of freedom, we included PMF calculations for center-to-center interactions with a 30° rotation applied to one of the platelet surfaces, as shown in Fig~\ref{fig:fig8}(b). The general trend remains the same, and hydrated layers appear at similar d-spacing. However, the barrier energy from 0-W to 1-W has increased in Ca- and Cu-MMT, possibly due to the more pronounced effect of hydrogen bonding at the edges. The difference in PMF profiles before and after rotation, especially for the Cu-MMT structure, may be attributed to the specific interactions of Cu$^{2+}$ ions with the bridge oxygen atoms, leading to distinct binding behaviors that are sensitive to the orientation of the platelets. It is worth mentioning that while a 0-W state is included in our simulations for theoretical completeness, in practical scenarios, some level of hydration is always present. However, in vapor at very low relative humidity (RH) or high temperatures, low hydration states can indeed become more favorable.

In our study, we conducted PMF calculations for interactions involving two edge-to-edge platelets according to Fig~\ref{fig:fig8}(c). This type of interaction does not exhibit significant dependence on the cations present in the environment. While the local minima in edge-to-edge interactions are not as pronounced as those observed in center-to-center interactions, we observe an oscillation pattern in the PMF profiles. This pattern is attributed to the presence of water layers positioned between the two platelets.

\begin{figure}[htbp]
    \centering
    \includegraphics[width=0.8\textwidth]{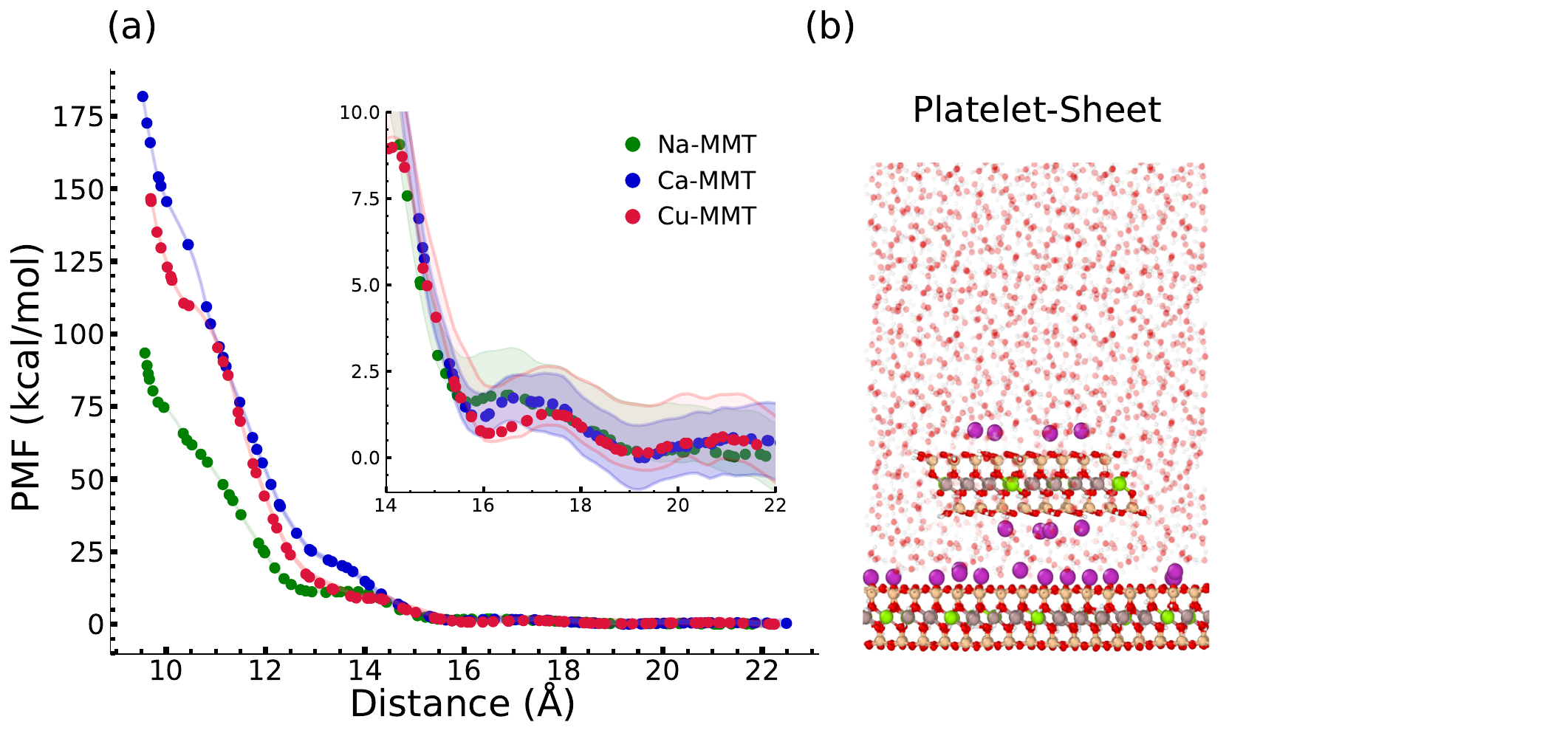}
    \caption{Potentials of mean force (PMF) for center-to-center configuration between one periodic sheet and one small platelet of Na-, Ca- and Cu-MMT. (a) The PMFs, represented by dots, are averaged, and the shaded colors indicate the variability obtained from 100 bootstrap samples. The zoomed-in graph reveals a well-defined third water minimum. (b) Corresponding structure for Na-MMT is shown.
}
    \label{fig:fig9}
\end{figure}

Another assessment involved studying the interaction between a single small platelet and an infinite (periodic) sheet to replicate the interaction between small and significantly larger platelets. The resulting PMF and an illustration of the system are presented in Fig~\ref{fig:fig9}. The average PMF curves share similarities with those of individual platelets. However, the initial energy barrier for the transition from 0-W to 1-W is lower compared to that observed for small platelets. This discrepancy can be attributed to the reduced contribution of hydrogen bonds in these interactions. The interaction energies exhibit oscillatory patterns as a consequence of the hydration forces that come into play when increasing the size of a single MMT layer, while also decreasing the influence of hydrogen bonds.

\begin{figure*}[htbp]
    \centering
    \includegraphics[width=1\textwidth]{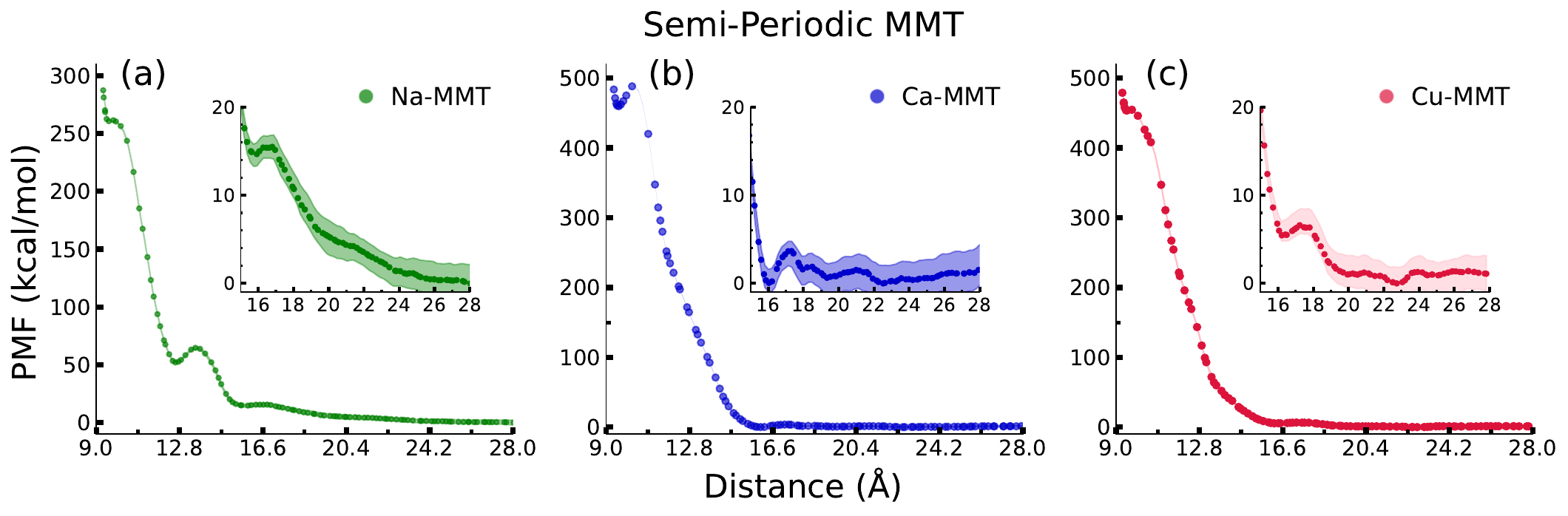}
    \caption{Potentials of mean force (PMF) for center-to-center orientation of two semi-periodic (a) Na-, (b) Ca- and (c) Cu-MMT. The PMF results, represented by dots, are averaged, and the shaded colors indicate the variability obtained from 100 bootstrap samples. Interestingly, the zoomed-in graph shows a higher slope for the Na-MMT interaction compared to Ca and Cu-MMT.
    }
    \label{fig:fig10}
\end{figure*}

The latest set of calculated PMF profiles, as depicted in Fig~\ref{fig:fig10}, pertains to the interaction between two semi-infinite (semi-periodic) sheets of MMT. This system effectively mimics the behavior of very large platelets while considering the influence of hydrogen bonds at the edges of these sheets, providing valuable insights into the interaction dynamics at this larger scale. As illustrated in Fig~\ref{fig:fig10}, the overall shape of the PMF curves exhibits similarities with those observed for smaller platelets. 
One important feature, as revealed by the zoomed-in PMF graphs in Fig~\ref{fig:fig10}, is the distinct behavior regarding water adsorption when the d-spacing exceeds approximately $\sim$ 21 Å. In this specific region, Na-MMT continuously adsorbs water without encountering any significant energy barriers. It appears to transition beyond the crystalline region, a phenomenon that aligns with a previous study on Arizona Na-MMT, which demonstrated the absence of forbidden layers in its structure \cite{fink1964x}. In contrast, Ca-MMT and Cu-MMT face relatively minor energy barriers in this d-spacing range and do not exhibit swelling. This disparity in behavior could be attributed to the presence of forbidden layers, as observed in the work of Meleshyn \textit{et al.} for Wyoming Na-MMT \cite{meleshyn2005gap}. The concept of forbidden layers arises from the chainlike structure formed by Na$^{+}$ ions and water molecules in the interlayer region. In the case of Wyoming Na-MMT, Na$^{+}$ ions form an inner complex, however; in Arizona MMT substitution occurs in the octahedral layer and Na$^{+}$ forms outer-sphere complex. This leads to the absence of forbidden layers in Arizona Na-MMT. 
However, it is plausible that Ca$^{2+}$ and Cu$^{2+}$ ions, due to their higher charges, form inner complexes with the Arizona MMT layers, resulting in the creation of forbidden layers. An experimental study using low-angle XRD has demonstrated that Ca-Wyoming, Ca-Otay bentonites,
and Ca-hectorite undergo a stepwise expansion, ultimately expansion stops at d-spacinng of 19.6 Å for Ca-Wyoming and 22 Å for other two clays \cite{fink1964x}. These findings support the presence of forbidden layers in Ca-MMT.

\subsubsection{Swelling pressure} 
In our study, we aimed to quantify and compare the swelling pressures induced by Cu-MMT systems with those of Na-MMT and Ca-MMT. To the best of our knowledge, limited information is available regarding the swelling pressures of Cu-MMT.

We employed the slopes derived from average PMFs to interpolate swelling pressures of Na-, Ca-, and Cu-MMT. The derivatives were obtained using cubic spline interpolation and the Savitzky-Golay filter. Additionally, we determined distances to establish dry densities, relying on the methodology introduced by Linlin Sun \textit{et al.} \cite{sun2015estimation}, which involved the conversion of d-spacing data into dry densities. As previously defined, these dry densities are calculated using the formula:
\[ \frac{m}{A \times d} \]
where \( m \) is the mass of the clay with ions, \( A \) is the area of the clay, and \( d \) is the d-spacing.
The PMF is equivalent to the grand potential in our system and can be expressed as:
\[
\text{PMF}(h) = \int_{h_0}^{h} P(h') \, dh' 
\]
At controlled pressure (\( P_{\text{control}} \)), the minimized potential is the osmotic potential (\( \lambda \)):
\[
\lambda = \text{Grand Potential} + P_{\text{control}} \cdot h
\]

In thermodynamic equilibrium, this osmotic potential is minimized. The difference in osmotic potential between two states can be expressed as:
\[
\lambda - \lambda_0 = \int_{h_0}^{h} (P - P_{\text{control}}) \, dh
\]
The value of \( P_{\text{control}} \) that provides the transition point is found where this integral equals zero. This indicates that the areas under and above the controlled pressure line on the swelling pressure curve are equal, ensuring phase equilibrium and indicating a phase transition \cite{brochard2017nanoscale, brochard2021swelling}.

\begin{figure}[htbp]
    \centering
    \includegraphics[width=1\textwidth]{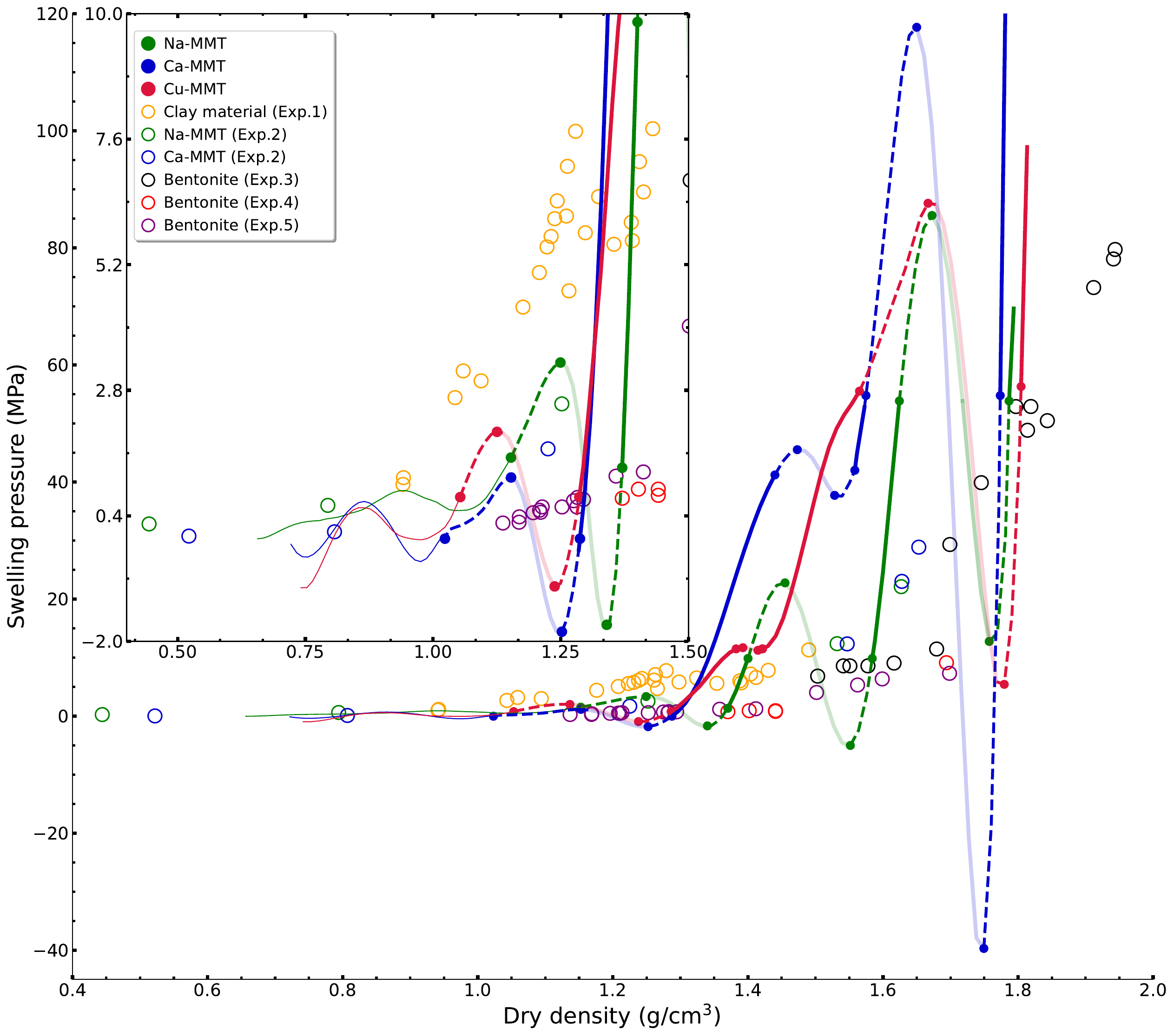}
    \caption{Comparison of simulated swelling pressures of Na-, Ca-, and Cu-MMT with experimental results as a function of dry density \cite{kumpulainen2011mineralogical, karnland2006mineralogy, bucher1984quelldruck, sitz1997materialuntersuchungen, agus2008method} Solid lines represent stable regions, dashed lines represent metastable regions, and transparent lines represent unstable regions. Filled markers and lines indicate our simulation results, while empty circles represent experimental data. Positive swelling pressures indicate layer expansion due to repulsive forces, while negative pressures indicate contraction due to attractive forces.}
    
    \label{fig:fig11}
\end{figure}

The obtained swelling pressures from our simulations are visually presented in Fig~\ref{fig:fig11}, marked with filled circle symbols and lines. Solid lines indicate stable regions, dashed lines indicate metastable regions and transparent lines indicate unstable regions. The points show the transitions, which are obtained using the explained equations.

The overall trend suggests that the stepwise stable regions can be described by two confining exponential equations. This is consistent with patterns observed in previous consistent with patterns observed in previous experimental and simulation studies with pure water and saline solutions \cite{sun2015estimation, akinwunmi2020swelling, akinwunmi2019influence, agus2008method}. 

To validate our simulation results, we compared them with data from experimental studies. In Fig~\ref{fig:fig11}, we include swelling pressure for 24 different bentonites, each with different compositions or exchangeable cations (Clay materials (Exp.1))\cite{kumpulainen2011mineralogical}. We also reference the swelling pressure data for Wyoming MX-80 bentonite, dominated by either Ca (Ca-MMT (Exp.2)) or Na (Na-MMT (Exp.2)) \cite{karnland2006mineralogy}. Additionally, we present measurements of swelling pressure in bentonite (Exp.3) containing 50-60\% MMT with Ca$^{2+}$ and Mg$^{2+}$ exchangeable cations \cite{agus2008method}, and compared them with Bucher \textit{et al.} and Sitz \textit{et al.} work (Bentonite (Exp.4-5)) \cite{bucher1984quelldruck, sitz1997materialuntersuchungen}.

For the entire range of densities, the stability trend of the swelling pressure for all structures lies between two exponential curves. For Na-MMT, the lower exponential curve aligns well with the experimental results, indicated by the empty circled markers \cite{kumpulainen2011mineralogical, karnland2006mineralogy, bucher1984quelldruck, sitz1997materialuntersuchungen, agus2008method}. This alignment suggests that our model accurately captures the swelling behavior of Na-MMT across this density range. Similarly, for Cu-MMT, the stability trend follows the same pattern. For Ca-MMT, the lower bound of stability aligns with the experimental results for both low and high densities, showing good agreement with experimental data.

At lower densities, below 1.3 g/cm³, Na-MMT exhibits relatively high swelling pressures compared to Ca-MMT and Cu-MMT. As the density increases from 1.3 to 1.65 g/cm³, the swelling pressure of Ca-MMT begins to surpass that of Na-MMT and Cu-MMT. Cu-MMT generally displays intermediate swelling pressures between Na-MMT and Ca-MMT across this density range.

It is important to note that while the lower bound of Na-MMT's swelling pressure aligns with experimental data, the trend observed for Ca-MMT, particularly in the density range of 1.35 to 1.65 g/cm³, where it surpasses the swelling pressure of Na-MMT, contrasts with experimental observations that typically show Na-MMT swelling more than Ca-MMT at these densities. This pattern, however, has been observed in other molecular dynamics simulation studies as well \cite{sun2015estimation, akinwunmi2020swelling}. The deviation from experimental results could be due to several factors, including the limitations of the force field used in the simulations, the assumptions made regarding the ionic environment, and the simplifications inherent in the model.

The overestimation in our calculations could be attributed to the idealized nature of our simulated models, which consist of two perfectly stacked MMT clays. In practice, Effective Montmorillonite Dry Density (EMDD) is used to account for factors influencing swelling mechanisms, and it is typically lower than the actual dry density (DD). Our comparison with EMDD-adjusted experimental data suggests that while our results are higher, this discrepancy can be attributed to differences in model assumptions and environmental conditions in simulations versus experiments. For instance, the percentage of water used in experiments versus simulations can influence the results, and experimental conditions are highly sensitive to environmental variables such as temperature, humidity, and sample preparation methods. Additionally, the charges on bentonite clay particles can affect the interaction with cations and contribute to discrepancies between simulated and experimental results. Further investigation and refinement of the simulation parameters may be needed to achieve better agreement with experimental data.

\subsection{\label{sec:citeref}Mobility of interlayer cations}
The PMF calculations for the semi-periodic MMTs identified approximate local minima for each structure, allowing us to track the trajectories of interlayer ions and subsequently compute their mean square displacements (MSD). After allowing sufficient time for the system to equilibrate, ensuring the amount of water between semi-infinite sheets and the forces reached equilibration, we replaced the semi-infinite platelets with the infinite platelets and measured the MSD based on the average MSD for ions in two interlayer regions.

\begin{figure}[htbp]
    \centering
    \includegraphics[width=1\textwidth]{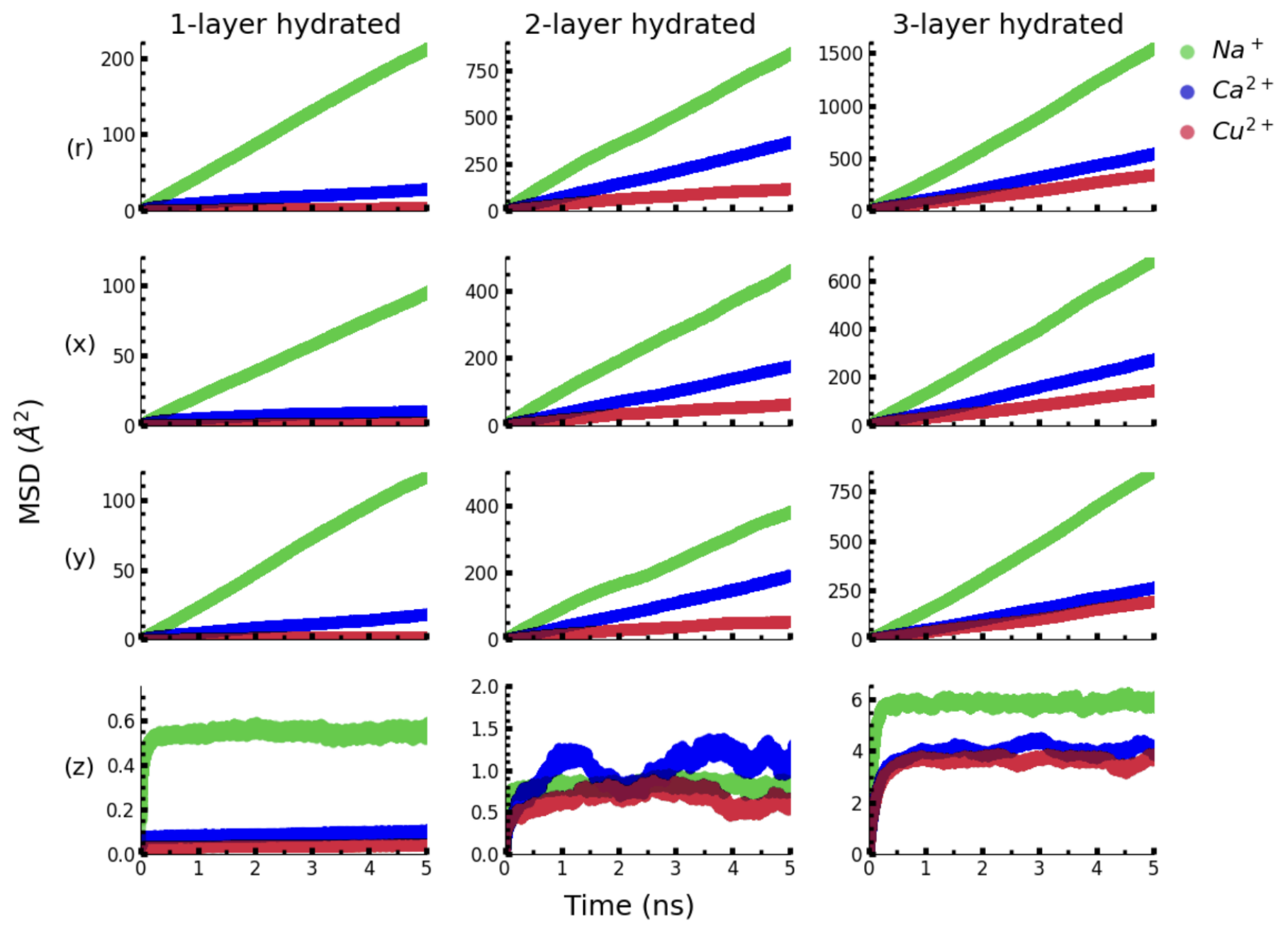}
    \caption{Mean square displacement (MSD) calculations for interlayer Na$^{+}$, Ca$^{2+}$, and Cu$^{2+}$ ions in one, two, and three hydrated layers. Columns represent the number of hydrated layers: first column (1 layer), second column (2 layers), and third column (3 layers). Rows display total MSD, as well as x, y, and z components of the MSD.
    }
    \label{fig:fig12}
\end{figure}

The MSD graphs presented in Fig~\ref{fig:fig12} depict the mobility of ions for three different water content scenarios: 1-W, 2-W, and 3-W layers. These graphs show averaged MSD calculations derived from 50 MSD runs over a 5 ns simulation period.

Across all scenarios, Na$^+$ ions exhibited higher mobility than the other ions. We quantified ion self-diffusion coefficients (D) and reported them with uncertainties in the form of standard errors, as detailed in Table~\ref{tab:table3}. For the diffusion calculations, we excluded the last 0.1 ns interval to avoid inaccuracies.

The obtained diffusion coefficients revealed that, although an increase in the number of hydrated layers led to a slight initial increase in diffusion in the Z-direction (perpendicular to the surface of MMTs) for all ions, the curves ultimately saturated. This saturation indicates that there was no significant diffusion in this direction across all cases. Therefore, we considered the total diffusion in the XY plane.

The self-diffusion coefficients of Na$^+$ ions in the 1-, 2-, and 3-W scenarios are consistent with prior literature, aligning closely with values of 0.53-1.09, 3.6-6.5 and 5.3-7.9 $\times 10^{-10} \, \text{m}^2/\text{s}$ reported at temperatures of 298 K and 323 K \cite{holmboe2014molecular}.

Similarly, the self-diffusion for Ca$^{2+}$ ions closely match findings from previous MD simulations involving highly charged MMT structures of -0.75 e per unit cell \cite{greathouse2016molecular}. This study reported self-diffusion coefficients of 0.6, 1.2, and $2.2 \times 10^{-10} \, \text{m}^2/\text{s}$ for 1-, 2-, and 3-W states, respectively, which is relatively consistent with our observations for MMT containing 0.67 e per unit cell.
Ca$^{2+}$ ions displayed lower diffusion compared to Na$^+$ ions, a consistent trend attributed to the more stable coordination shell of Ca$^{2+}$ as reported in prior research \cite{zhang2014hydration}. 

Interestingly, the diffusion pattern of Cu$^{2+}$ ions is lower than that of Ca$^{2+}$ ions in all directions and for all three hydration layers. However, when compared to Na$^{+}$ ions, the diffusion behavior of Cu$^{2+}$ ions is more similar to that of Ca$^{2+}$ ions. This suggests that, despite Cu$^{2+}$ ions having lower mobility than Ca$^{2+}$ ions, they share more comparable mobility characteristics with Ca$^{2+}$ ions than with Na$^{+}$ ions in the context of different hydration layers. The higher diffusion rate of Na$^{+}$ ions is due to their single positive charge, which results in weaker interactions with water molecules compared to the divalent Ca$^{2+}$ and Cu$^{2+}$ ions. Ca$^{2+}$ and Cu$^{2+}$ have similar diffusion behaviors because of their similar charges, but the slightly lower diffusion rate of Cu$^{2+}$ can be attributed to its somewhat smaller hydrated ionic radius, which leads to stronger interactions with the surrounding water molecules, thus reducing its mobility.


\begin{table*}[htbp]
    \caption{Comparison of self-diffusion coefficients (in 10$^{-10}$ m$^2$/s) for Cu$^{2+}$, Ca$^{2+}$, and Na$^{+}$ at three different hydration levels in the interlayer regions.}

    \scriptsize
        \begin{tabularx}{\textwidth}{l X X X}
            \hline
            Layers of hydration &   D$_{\text{Na$^{+}$}}$  & D$_{\text{Ca$^{2+}$}}$ & D$_{\text{Cu$^{2+}$}}$  \\ 
            \hline
            1-layer hydrated &   &  &   \\
            X-direction & 0.99 $\pm$ 6.25 $\times 10^{-5}$  & 0.09 $\pm$ 1.94 $\times 10^{-5}$ & 1.61 $\times 10^{-2}$ $\pm$ 1.62 $\times 10^{-6}$  \\
            Y-direction & 1.09 $\pm$ 6.57 $\times 10^{-5}$  & 0.16 $\pm$ 3.39 $\times 10^{-5}$ & 1.11 $\times 10^{-2}$ $\pm$ 6.05 $\times 10^{-6}$  \\
            Total & 1.04 $\pm$ 3.57 $\times 10^{-5}$  & 0.13 $\pm$ 2.19 $\times 10^{-5}$ & 1.32 $\times 10^{-2}$ $\pm$ 3.36 $\times 10^{-6}$  \\
            \hline
            2-layer hydrated &   &  &   \\
            X-direction & 4.98 $\pm$ 4.56 $\times 10^{-4}$  & 1.98 $\pm$ 3.50 $\times 10^{-4}$ & 0.53 $\pm$ 5.47 $\times 10^{-5}$  \\
            Y-direction & 3.90 $\pm$ 2.68 $\times 10^{-4}$  & 1.38 $\pm$ 5.14 $\times 10^{-4}$ & 0.42 $\pm$ 7.98 $\times 10^{-5}$ \\

            Total & 4.44 $\pm$ 3.2 $\times 10^{-4}$  & 1.58 $\pm$ 1.2 $\times 10^{-4}$ & 0.49 $\pm$ 6.93 $\times 10^{-5}$ \\
            \hline
            3-layer hydrated &   &  &   \\
            X-direction & 7.34 $\pm$ 3.34 $\times 10^{-4}$  & 2.31 $\pm$ 4.24 $\times 10^{-4}$ & 1.15 $\pm$ 2.33 $\times 10^{-4}$  \\
            Y-direction & 7.74 $\pm$ 5.64 $\times 10^{-4}$ & 2.32 $\pm$ 3.96 $\times 10^{-4}$ & 2.25 $\pm$ 4.15 $\times 10^{-4}$  \\
            Total & 7.53 $\pm$ 1.83 $\times 10^{-4}$  & 2.29 $\pm$ 3.79 $\times 10^{-4}$ & 1.72 $\pm$ 1.42 $\times 10^{-4}$  \\
            \hline
        \end{tabularx}
    \label{tab:table3}
\end{table*}

\section{Discussions and conclusions}
\label{sec:Discussions and conclusions}
In this work, we employed DFT calculations for Cu-MMT and extended the ClayFF force field to include Cu$^{2+}$ interactions, allowing for the accurate modeling of Cu-MMT systems. We examined the interaction energies between two MMT with three different cations including Na$^{+}$, Ca$^{2+}$, and Cu$^{2+}$. Our calculations suggest that, in pure water solution, the swelling pressure induced by Cu-MMT falls between that of Ca-MMT and Na-MMT, a significant observation with implications for DGR applications. Our study suggests that exerted swelling pressure caused by Cu-MMT on the container is  comparable to those of Ca- and Na-MMT.  Also, Cu-MMT exhibits substantial swelling capabilities similar to Na- and Ca-MMT, effectively hindering bacterial activities, which is crucial for the protection of the container and therefore the containment of used nuclear fuel. Of note, in a DGR system, the actual conditions might involve the presence of saline solutions, which could influence the Cu ion charge state, the swelling pressure induced by Cu-MMT and its implications for DGR applications.

Furthermore, our MSD studies demonstrated that Na$^+$ ions exhibit higher mobility than Ca$^{2+}$ and Cu$^{2+}$ ions, particularly in the XY plane. We observed that the diffusion rate increases with the hydration layer for all ions and in all directions. This observation indicates that additional water layers facilitate greater ion mobility, likely due to reduced interactions with the clay surface as hydration increases.

We found that the diffusion of Cu$^{2+}$ ions was lower than that of Ca$^{2+}$ ions in all directions and for all three hydration layers. This behavior is expected based on their different chemical properties. Our calculations showed that, in the event of copper corrosion from the container, these Cu$^{2+}$ ions can be captured by the surrounding bentonite.


In this research, we considered Cu$^{2+}$ as the charge state of copper ions in the clay systems. However, copper may exist in other charge states. This assumption is empowered by the results of energy profile analysis of more than 40 L-J potentials for Cu$^{2+}$ along the perpendicular direction of MMT, which two of the best results are presented in Fig~\ref{fig:fig2}. Despite testing a variety of potentials, we found that our L-J potential captures the well depth and convexity of the curvature around the equilibrium better than other tested L-J potentials, but it still does not perfectly match the DFT results. This discrepancy suggests that the charge of the copper ion may not be exactly 2+ in the presence of dry MMT. Furthermore, it is possible that the charge of the copper ion varies with its distance from MMT. In future, we will explore the charge distribution of copper ion using Bader charge analysis in dry and wet clay systems to gain deeper insight into these phenomena.

Overall, our findings provide a better understanding of the behavior of interlayer cations in clay systems. The extension of the ClayFF force field to include Cu$^{2+}$ interactions enhances its versatility and applicability in modeling a broader range of clay systems. In the future, we aim to use the PMFs in this study to develop coarse-grained mesoscales of hydrated Cu-MMT, Ca-MMT and Na-MMT and examine their mechanical properties. An interesting extension of this work would involve including a PMF calculation where the rotation of one platelet around the Z-axis is relaxed, allowing for the actual minimum PMF to be obtained. This could potentially alter the observed swelling pressures and provide a more accurate representation of the interaction characteristics.

\begin{acknowledgement}

This work is financially supported by the Nuclear Waste Management Organization, the Natural Science and Engineering Research Council (NSERC), and Mitacs. We also thank the Digital Research Alliance of Canada and Centre for Advanced Computing at Queen’s University for generous allocation of computer resources.

\end{acknowledgement}


\bibliography{achemso-demo}

\end{document}